\documentclass{article}

\usepackage{PRIMEarxiv}

\usepackage{xcolor}
\usepackage[utf8]{inputenc} 
\usepackage[T1]{fontenc}    
\usepackage{hyperref}       
\usepackage{url}            
\usepackage{booktabs}       
\usepackage{amsmath}
\usepackage{bbm}
\usepackage{amsfonts}
\usepackage{amssymb}
\usepackage{pifont}         
\usepackage{cleveref}  
\newcommand{\xmark}{\ding{55}}
\newcommand{\cmark}{\ding{51}}
\usepackage{algorithm}
\usepackage{algpseudocode}
\usepackage{amsfonts}       
\usepackage{nicefrac}       
\usepackage{chngpage}
\usepackage{calc}
\usepackage{listings}
\usepackage{xpatch}
\xpatchcmd{\mintinline}{\begingroup}{\begingroup}{}{}
\xpatchcmd{\minted}{\VerbatimEnvironment}{\VerbatimEnvironment}{}{}

\usepackage{microtype}      
\usepackage{float}
\usepackage{fancyhdr}       
\usepackage{graphicx}       
\usepackage{bbm}            
\usepackage{xcolor}
\definecolor{myred}{RGB}{145,78,97} 
\usepackage{makecell}       
\usepackage{caption}
\usepackage{subcaption}
\graphicspath{{media/}}     
\usepackage{tikz}
\usetikzlibrary{arrows.meta}
\usepackage{varwidth}
\usetikzlibrary{decorations.pathreplacing, math,shapes.misc}
\usepackage{hhline}
\usepackage{multirow}
\usepackage[export]{adjustbox}
\usepackage{hyperref}

\usepackage[]{apacite}
\usepackage{amsthm}
\theoremstyle{plain}

\title{A machine learning approach based on survival analysis for IBNR frequencies in non-life reserving}

\author{
  Munir Hiabu \\
  University of Copenhagen \\
  Copenhagen\\
  \texttt{mh@math.ku.dk} 
   \And
  Emil D. Hofman \\
  University of Copenhagen \\
  Copenhagen\\
  \texttt{edh@math.ku.dk} 
   \And
  Gabriele Pittarello \\
  Sapienza, Università di Roma\\
  Roma\\
  \texttt{gabriele.pittarello@uniroma1.it} 
}

\begin{document}
\maketitle

\begin{abstract}
We introduce new approaches for forecasting IBNR (Incurred But Not Reported) frequencies by leveraging individual claims data, which includes accident date, reporting delay, and possibly additional features for every reported claim. A key element of our proposal involves computing development factors, which may be influenced by both the accident date and other features. These development factors serve as the basis for predictions. While we assume close to continuous observations of accident date and reporting delay, the development factors can be expressed at any level of granularity, such as months, quarters, or year and predictions across different granularity levels exhibit coherence. The calculation of development factors relies on the estimation of a hazard function in reverse development time, and we present three distinct methods for estimating this function: the Cox proportional hazard model, a feed-forward neural network, and eXtreme gradient boosting. In all three cases, estimation is based on the same  partial likelihood that accommodates left truncation and ties in the data. 
While the first case is a semi-parametric model that assumes in parts  a log linear structure, the  two machine learning approaches only assume that the baseline and the other factors are multiplicatively separable.
Through an extensive simulation study and real-world data application, our approach demonstrates promising results. 
\end{abstract}

\section{Introduction}
\label{sec:introduction}

IBNR (Incurred But Not Reported) refers to outstanding claims for which the insurer is liable but which have not been reported at the time the reserve is calculated. Empirical studies on insurance markets show that in many lines of business the estimated cost of IBNR claims is the most important provision for the insurer, see for example \cite[p.~80]{friedland10}. As the number of IBNR claims is not known at the time the reserve is calculated, there is a strong actuarial argument in favour of methods that accurately predict the number of IBNR claims. 

In the seminal paper \citeA{miranda2013continuous}, the authors discuss how the common run-off triangles encountered in loss reserving in non-life insurance  can be understood within a  continuous framework where the goal is to estimate the distribution in the lower triangle.
Within this continuous chain-ladder framework, two lines of research emerged:
One aims to estimate the underlying density function and with this respect,
\citeA{lee15}, \citeA{lee17}, and \citeA{mammen21} generalize the initial model to account for 
seasonal effects, operation al time and calendar effects, respectively. 
In the other line of research, \cite{hiabu2016sample} establishes that the observation scheme of a continuous run-off triangle can be understood as right-truncation problem and that by reversing the development time statistical analysis can be conducted 
under a tractable  left-truncation setting. Building on this work,
\cite{hiabu17} discusses how the hazard function in reverse development time is related to the omnipresent development factors. \cite{hiabu2021smooth} extends the framework by allowing for accident date effects and other features via a multiplicative structure. \cite{bischofberger20} extends the prediction of claim frequencies to the prediction of claim payments. A simulation study comparing the two lines of research  can be found in \cite{bischofberger2019comparison}.

A major drawback of both lines of research so far is that they have not been very practical: Estimation is based on kernel smoothers
that without further adjustment do not cope well with the sharp patterns often seen in claim developments.
Additionally, it is not directly clear how to include categorical features.
In fact, with the exception of \cite{hiabu2021smooth} that allows for continuous features, none of the other work so far considers additional features, neither continuous nor categorical. A limiting factor is that if no structure is imposed  a priori, kernel smoother will necessarily suffer from the curse of dimensionality, i.e., exponentially deteriorating estimation performance with every continuous feature added.
Machine learning methods on the other hand have shown that they are capable of data-driven dimension reduction  and
making use of the underlying data structure in order to circumvent the curse of dimensionality. 

Our proposal extends the second stream of research by
proposing to maximize a partial likelihood function to estimate an accident date and other features dependent hazard function.
While any statistical machine learning method can be used to do so, in this paper we consider three  methods: the Cox proportional hazard model, a feed-forward neural network and eXtreme gradient boosting. 
In all of the three proposed methods, estimation will be based on the same  partial likelihood that accommodates left truncation and ties in the data.  While there is a growing number of survival analysis solutions being implemented for common machine learning methods, to the best of our knowledge, neither of the  publicly available  feed-forward neural network solution  nor eXtreme gradient boosting solutions  have implementations that can deal with left-truncated data and  ties in the data \cite{wiegrebe23}.

Hence, to make our proposal work, we will extend  current eXtreme gradient boosting and feed-forward neural networks solutions  such  that they  can handle 
the proposed likelihood function.
A particular focus in our proposal is the consistent transition between continuous and discrete objects, i.e., how to transform a hazard function into a development factor for run-off triangles.

To this end, we introduce the same approximation (managed by a parameter $\eta$, see also \citeA{pittarello25})
when handling ties in the baseline hazard and transforming the estimated hazard into a development factor.

We will use the estimated hazard function to calculate development factors which we will subsequently use for prediction. With this respect, our proposal is related to the approach of \cite{wuthrich18}
that uses a feed-forward neural network to estimate the development factors directly. However, we highlight three differences:
1) \cite{wuthrich18} employs a separate neural network for every development period; making it a) inherently a discrete in development time direction and b) computationally prohibitively expensive if a too granular development time grid is chosen. In contrast we only need to train a single neural network and assume (close to) continuous observations
in development time.
2) In our survival analysis approach, we do not have problems with zero entries in the cumulative run-off triangles. In \cite{wuthrich18}, the loss function fed into the neural network 
entails dividing by each cumulative entry, see equation (3.1) in that paper. While zero entries are 
not common in typical run-off triangles, zero entries are expected to happen often if the features take too many different values resulting in many sparse triangles. In particular, the approach of \cite{wuthrich18} does not allow for continuous features. To circumvent the problem of some zero entries with discrete or categorical features, \cite{wuthrich18} proposes some adhoc method (based on rather strong assumptions) that ignores features in those entries.  In contrast to that, in our  approach we do not have those problems with zero entries and are able to estimate  a conditional hazard function which is  possibly dependent on high dimensional feature information. 3) In \cite{wuthrich18}, it is assumed that the development factors, as is the case in chain ladder, do not vary with accident dates.
In our approach, accident date is allowed to alter the development factors, hence providing a more flexible approach.

In a recent preprint, \cite{calcetero2023claim} aim to estimate the same hazard function as we propose to estimate.
However, they only apply the Cox-proportional hazard model and not further machine learning methods as we do. Furthermore, they use an inverse probability weighting
for prediction. In contrast, we propose to transform the hazard function into development factors which are subsequently used for prediction. While one may expect that both Cox-proportional hazard models will produce similar predictions, one advantage of our proposal is the ease of comparison with standard reserving methods based on development factors; see the next section.

We organize our manuscript as follows. We conclude the introduction of this manuscript in \cref{ss:overview}, where we show an overview of the models development factors output based on eXtreme gradient boosting on a simulated dataset. In \cref{s:modeling} we introduce the continuous time framework and the hazard function we wish to estimate, while 
\cref{s:estimation} introduces various estimation strategies.

\Cref{s:modelingtheclaimsdev}  establishes the connection between the  estimated hazard functions and development factors.
 In \cref{sec:evaluation} we discuss the performance measures that we will use in the empirical analyses to select and compare our models. In \cref{sec:dataapplicationsim} we will challenge our models on several data sets simulated from $5$ different scenarios. 
 In \cref{sec:dataapplicationrd} we present an application on a real dataset from a Danish insurer. The dataset is not publicly available. We conclude the manuscript with general remarks on our approach.

\subsection{Overview of the main model output}
\label{ss:overview}

In recent years, there has been a growing number of papers developing reserving methods based on individual data; also known as micro-level reserving or granular reserving.
See e.g.  
\cite{fung2022fitting, crevecoeur2022hierarchical,lopez21, delong22, michaelides23}
for some recent contributions  and references therein. 

The complexity of micro-level models makes their practical implementation challenging. Indeed, while the insurance industry is experimenting with innovative data-driven projects, it is still a long way from implementing individual reserving models in practice \cite{brown23}. A recent study by the Actuarial Studies in Non life Insurance (ASTIN) section of the International Actuarial Association (IAA) shows that companies tend to rely on simple algorithms with basic software implementation \cite{astin16}. This is due to the difficulty in obtaining data of sufficient quality for sophisticated data analysis and the preference for licensed software that does not include novelties from the literature. For these reasons, the chain-ladder remains the usual approach to reserving.

In  this brief excerpt we aim to show how our model output can be  visualized in a way familiar to reserving actuaries.
It also  allows a seamless comparison with estimations obtained through the chain-ladder approach.
To this end, we show a preview of the results that we obtained on a simulated data set coming from scenario Delta,
as will be introduced in  \cref{sec:dataapplicationsim}. In our simulation study, a claim can be of \texttt{claim\_type 0} or \texttt{claim\_type 1}.
In scenario Delta,  we consider a seasonality effect such that the development factors dependent on the accident date.

This could occur in a real world setting with an increased work load during winter for certain claim types, or a decreased workforce during the summer holidays. 

\begin{figure}[H]
    \centering 
\begin{subfigure}{0.35\linewidth}
  \includegraphics[width=0.8\linewidth]{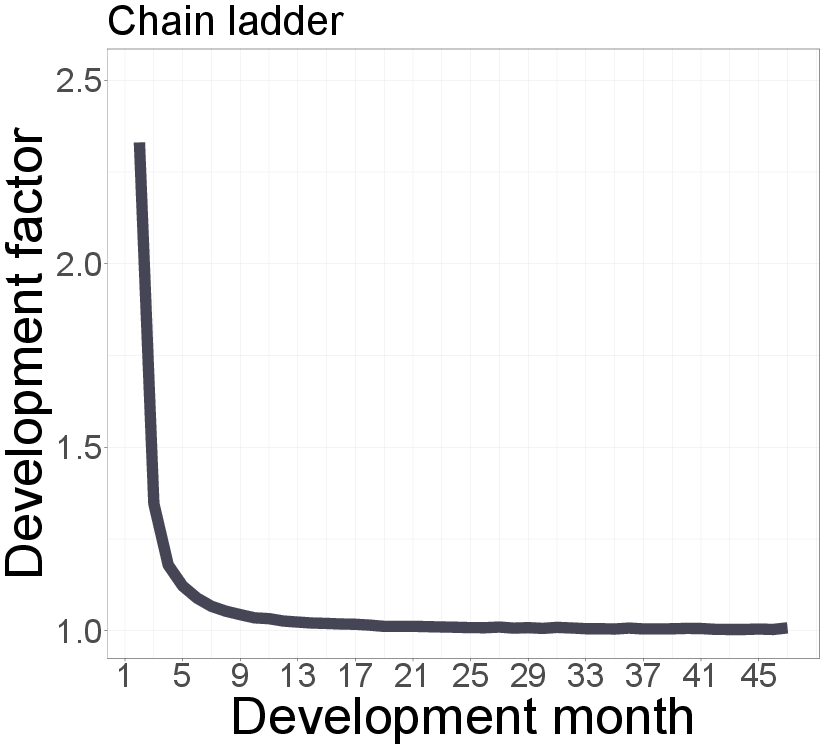}
  \caption{}
  \label{fig:clq}
\end{subfigure}\hfil
\begin{subfigure}{0.35\linewidth}
  \includegraphics[width=0.8\linewidth]{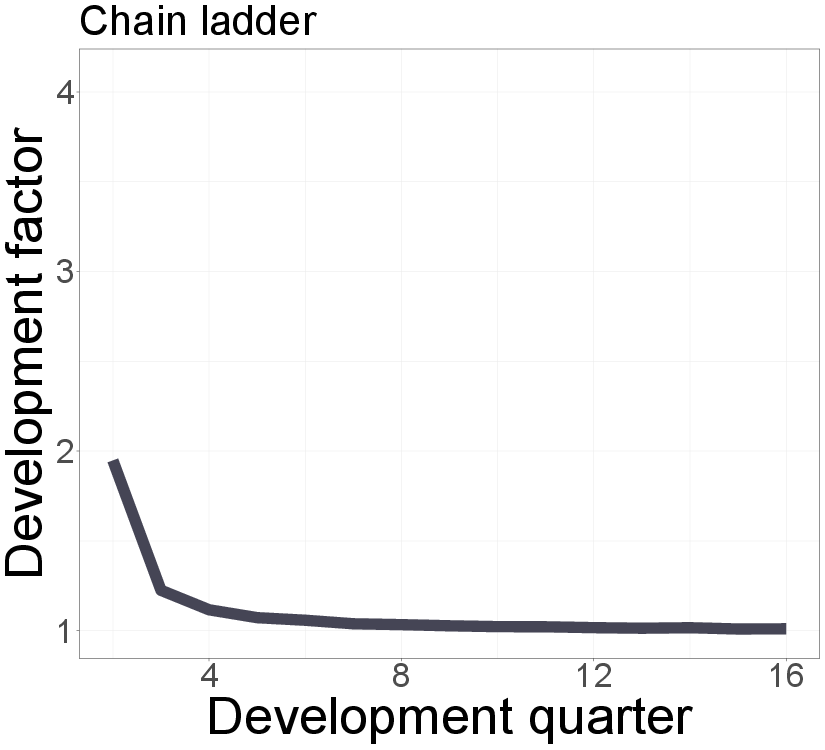}
  \caption{}
  \label{fig:cly}
\end{subfigure}\hfil
\medskip
\begin{subfigure}{0.32\linewidth}
  \includegraphics[width=0.8\linewidth]{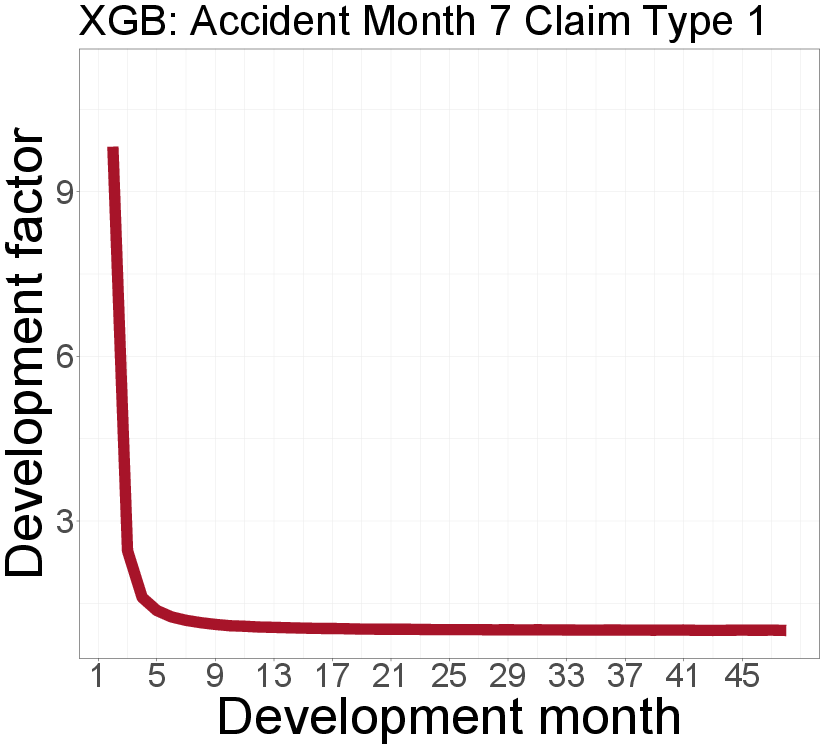}
  \caption{}
  \label{fig:4}
\end{subfigure}\hfil 
\begin{subfigure}{0.32\linewidth}
  \includegraphics[width=0.8\linewidth]{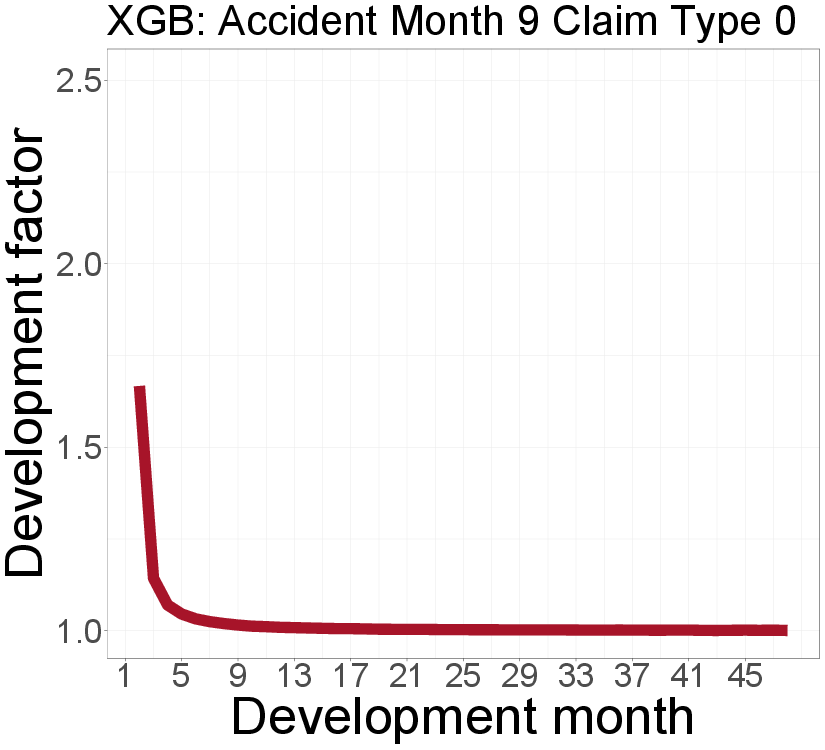}
  \caption{}
  \label{fig:5}
\end{subfigure}\hfil 
\begin{subfigure}{0.32\linewidth}
  \includegraphics[width=0.8\linewidth]{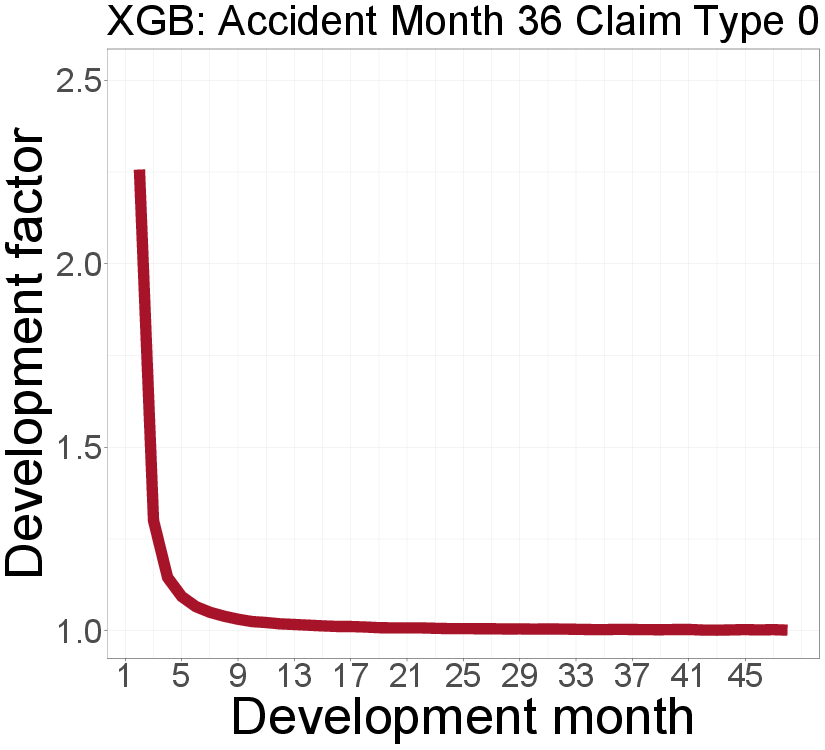}
  \caption{}
  \label{fig:6}
\end{subfigure}

\begin{subfigure}{0.32\linewidth}
  \includegraphics[width=0.8\linewidth]{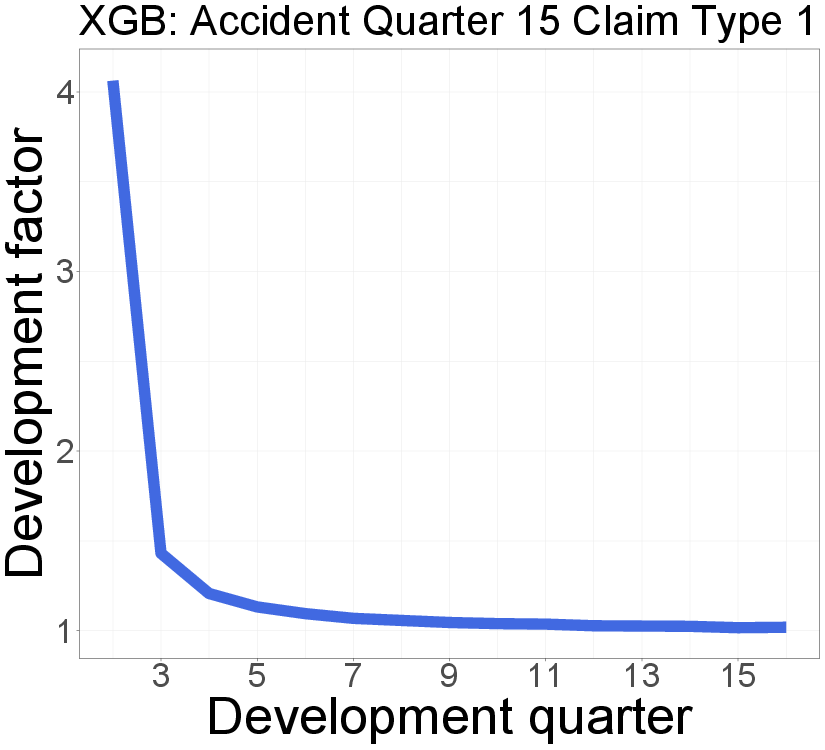}
  \caption{}
  \label{fig:1}
\end{subfigure}\hfil 
\begin{subfigure}{0.32\linewidth}
  \includegraphics[width=0.8\linewidth]{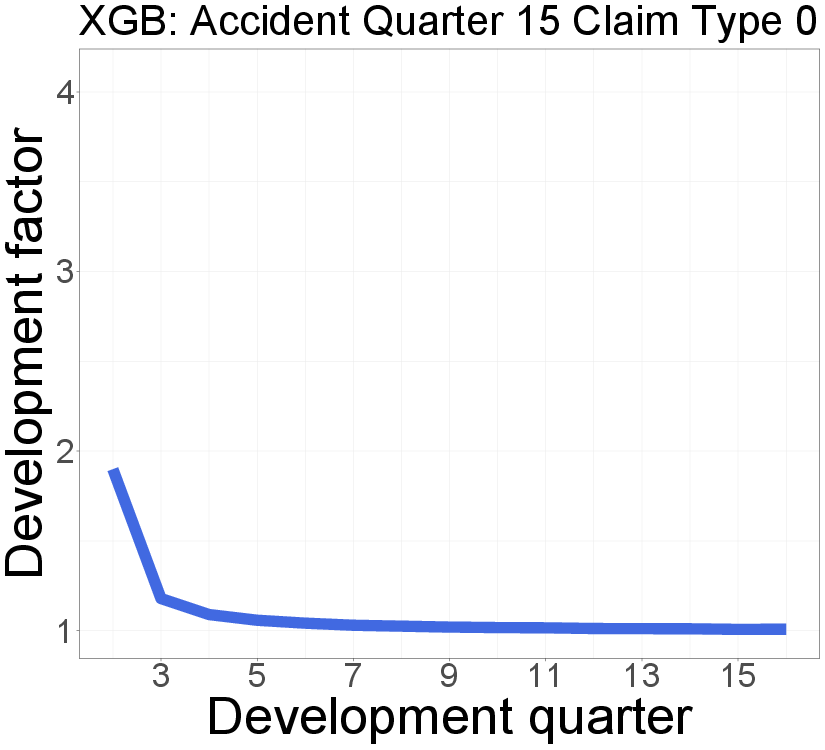}
  \caption{}
  \label{fig:2}
\end{subfigure}\hfil 
\begin{subfigure}{0.32\linewidth}
  \includegraphics[width=0.8\linewidth]{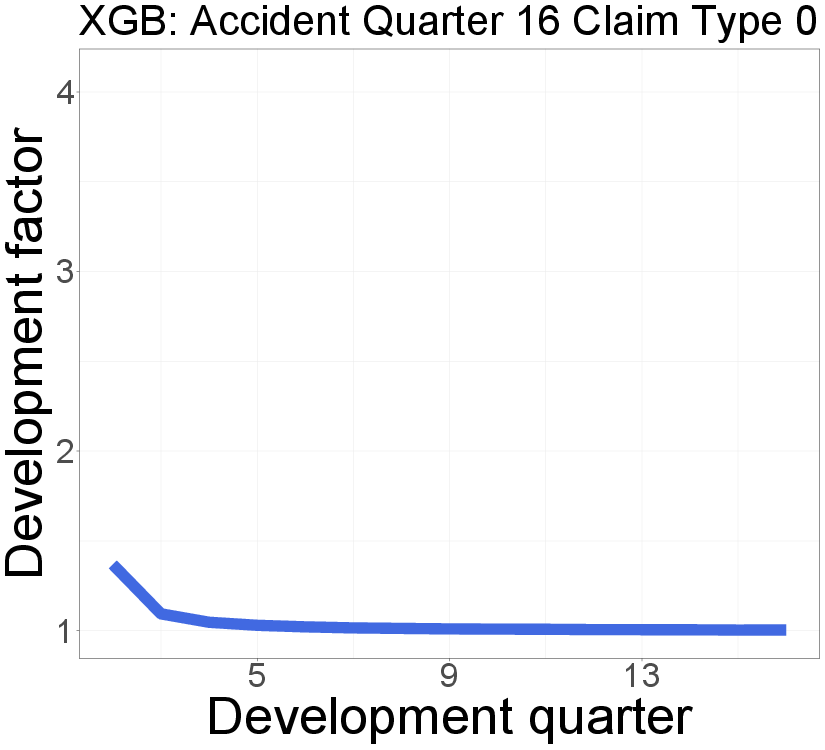}
  \caption{}
  \label{fig:3}
\end{subfigure}
\caption{The first row, shows the chain-ladder development factors fitted on monthly data (\autoref{fig:clq}), and quarterly data (\autoref{fig:cly}). The development factors do not consider additional features we have at our disposal.
The second and third row show an eXtreme gradient boosting output from our model proposal that depends on accident date and claim type.
The second row, shows monthly development factors for the feature combinations  Accident Month 7 and \texttt{Claim\_type 1}  (left panel),   Accident Month 9 and \texttt{Claim\_type 0} (center panel) and   Accident Month 36 and \texttt{Claim\_type 1}  (right panel). The third row shows quarterly development factors for the feature combinations Accident Quarter  15  and \texttt{Claim\_type 1}   (left panel),   Accident Quarter 15 and \texttt{Claim\_type 0} (center panel) and   Accident Quarter 16 and  \texttt{Claim\_type 0}  (right panel). 
}
\label{fig:images}
\end{figure}

An important feature of our proposal is that while inputs are assumed continuous, development factors and predictions can be provided on chosen granularity levels.
Oppositely to the chain-ladder model the predicted development factors will depend on accident date and additional feature information; here claim type being equal zero or one.
 In \cref{fig:images} we select one simulation and compare the chain-ladder development factors to our output development factors for different granularities (monthly and yearly). 
 The first row shows chain-ladder's development factors.
 The second row shows the output on a monthly level from an eXtreme gradient boosting fit for three different feature combinations of accident date and claim type. The three plots can be compared to the monthly development factors obtained from chain-ladder (first row left panel).
 The third row is the analogue to the second row, but this time a quarterly aggregation has been chosen and it can be compared to the quarterly development factors obtained from chain-ladder (first row right panel).
\newpage
 
\section{Modeling}
\label{s:modeling}

At a cut-off-date $\mathcal{U}$, we have observed $n$ claim reports. For each claim $i$ with $i=1,\dots,n$, we are given the accident date $U_i$ and the time delay from accident until report, $T_i$.
The variable $T_i$ is assumed continuous 
while $U_i$ is assumed  discrete and only takes values 
of the form $U_i=k\delta$, $\delta>0$; $k=0,1,\dots, K$.
For the latter  assumption to not cause a too big bias, one will need $\delta$
to be rather small. In our applications $\delta$ equaling one day worked well.
Later, when discussing estimation,  we will additionally assume that only discrete
approximations of $T_i$ are recorded.
The variables $T_i$ and $U_i$ and encoded such that $0\leq U_i \leq \mathcal U=K\delta; 0\leq T_i \leq \mathcal T$, $\mathcal U\geq \mathcal T$, i.e, $\mathcal T$ is the maximum delay and we require that the range of observable accident dates is larger than the maximum possible delay.

For each individual we have a set of $p$ measurements, i.e., features ${X}_i \in \mathbb{R}^p$. We assume that all reportings are independent.

Direct inference on $T_i$ may lead to sampling bias, as $T_i$ is observed only if the report happens before the cut-off date $\mathcal{U}$:
\[
 T_{i}\leq \mathcal{U} -U_i,
\]
which is a right-truncation problem. A solution to the right-truncation problem is to look at time in reversed direction leading to a tractable left-truncation problem \cite{ware76}. 
Concretely, we target $\mathcal T - T_i$ instead of $T_i$ such that the right truncation problem is now  a left truncation problem ($\mathcal T - T_i \geq U_i - \mathcal U + \mathcal T$) with truncation variable $(U_i- \mathcal U + \mathcal T)$.
We consider the reversed development time counting processes
$$N_i(t)= I(t \geq \mathcal{T} - T_{i}),$$
each with respect to the filtration $\mathcal F_{it}=\sigma \left( \bigg\{ \mathcal T - T_i \leq s:\ s\leq t\bigg\} \cup \bigg\{    U_i \bigg\} \cup \bigg\{X_i\bigg\} \cup \mathcal N\right)$, satisfying the \textit{usual conditions} \cite[p.~60]{andersen12}, and where $\mathcal N$ is the set of all zero probability events.

\subsection{The intensity process}
\label{sec:intensityprocess}

Assuming that the intensity $\lambda_{i}$ of the counting process exists and is piecewise continuous, we have 

\begin{equation}
\label{eq:hazard}
\lambda_{i}(\mathcal T -t|U_i, {X}_i): =\lim_{h \downarrow 0} h^{-1} E\left[  N_i\left\{(\mathcal{T}-t+h)-\right\}- N_i\left\{(\mathcal{T}-t)-\right\}|\ \mathcal F_{i,(\mathcal{T}-t)-}\right]=\alpha(t|U_i, {X}_i)Y_i(t),    
\end{equation}

where 
\begin{align*}
    \alpha(t|{u},{x})&=\lim_{h \downarrow 0} h^{-1}{P}\left(T_i\in(t-h,t]|\ Y_i(t)=1, {X_i}=x, U_i = u\right),\\
   Y_i(t)&=I(T_i \leq t <  \mathcal U -U_i ).
\end{align*}
Note that $Y_i(t)$ and $\alpha(t|U_i,X_i)$ correspond to
the intensity $\lambda_{i}(\mathcal T -t|U_i, {X}_i)$,
meaning  the development time input for $Y_i(t)$ and $\alpha(t|U_i,X_i)$  is not in reversed direction.
The structure, $\lambda_{i}(\mathcal T -t|U_i, {X}_i)=\alpha(t|U_i, {X}_i)Y_i(t)$, is called Aalen's multiplicative intensity model 
\cite{aalen78}, and enables nonparametric estimation and inference 
on the deterministic hazard function $\alpha(t|u,x)$.
We propose to model the hazard function as 
\begin{equation}
\label{eq:proportionalhazard}
\alpha(t|u,x) = \alpha_0(t)e^{\phi \left( u, x\right)}, 
\end{equation}
where $\alpha_0(t)$ is called the baseline hazard and $e^{\phi \left( u, x\right)} $ is the risk score; a component that depends on the features ${X}_i$ and  the accident period $U_i$. 
By assuming that the effects of $t$ and $u$ are multiplicatively  separated allows us to have predictions for $\alpha$ in the lower triangle,  $\{(t,u): t>\mathcal U-u\}$,
without extrapolation. In the next sections, we will discuss how to specify and model the log-risk function $\phi \left( u, x\right)$. We will consider three different models: the Cox proportional hazard model with splines \cite<COX,>{gray92}, neural networks \cite<NN,>{katzman18} and  eXtreme gradient boosting \cite<XGB,>{chen16}.

\begin{itemize}
    \item In COX, the log-risk function is assumed to be linear, $\phi \left( u, x\right)=\theta^T x + \theta_u u$, with $\theta \in \mathbb{R}^p$ and $\theta_u \in \mathbb{R}$. In this paper we will consider the more general log-risk function that includes splines for modeling continuous features.
    \item In NN, $\phi \left( u, x\right)$ a feed-forward neural network.
    \item In XGB, $\phi \left( u, x\right)$ is an ensemble of decision trees, i.e., functions piecewise constant on rectangles.
\end{itemize}

\section{Estimation}
\label{s:estimation}
\subsection{Ties in the data}

While we assume that $T_i$  is continuous, in practice, many of the  recorded reporting delays can have the same value, especially when the data records are not very granular. Even daily records can often have multiple occurrences with the same reporting delay.
To formalize this,
we partition the interval $[0,\mathcal T]$ into $J$ sub-intervals of equal length  $\delta$ with
boundaries 

$$0=t^{\left(0\right)}  <  \ldots < t^{\left(j\right)} < \ldots   < t^{(J)}=\mathcal T,
$$

such that for $T_i \in [t^{(j)},t^{(j+1)})$, the recorded value is $\text{rec}(T_i)=t^{(j)}$.
Additionally, for $T_i=\mathcal T$ the recorded value is $t^{(J)}$.
Our proposed estimators will only depend on $\text{rec}(T_i)$ and not $T_i$.

For $j=0,\ldots, J-1$,  we define the exposure set

$$ 
\mathcal{R}(t^{(j)})=
\left\{ i \in \left\{1, \ldots, n\right\}: T_i <  t^{(j+1)}; \   U_i < \mathcal{U} - t^{(j+1) }   \right\},
$$

and the occurrence set
$$
\mathcal{O}(t^{(j)}) =  \left\{ i \in \left\{1, \ldots, n\right\}:  rec(T_i)=t^{(j)} \right\},
$$
while we indicate with $O_j=\# \mathcal{O}(t^{(j)})$ the cardinality of the set $\mathcal{O}(t^{(j)})$.

To estimate the log-risk function $\phi$, we will first formulate the partial likelihood in a general form and then we will specify the log-risk function. The specifications we use (COX, NN, XGB) were briefly introduced in \cref{sec:intensityprocess}.

\subsection{Partial likelihood}
If $T_i$ would have been observed one could  directly use the partial likelihood defined as
\begin{align*}
\mathcal{L}^{\textrm{exact}}&= \prod_{i=1}^n \frac{e^{\phi(U_i, X_i)}}{\sum_{l=1}^n Y_l(T_i) e^{\phi(U_l, X_l)}}\\
&= \prod_{i=1}^n \frac{e^{\phi(U_i, X_i)}}{\sum_{l: T_l \leq T_i; U_l< \mathcal U-T_i} e^{\phi(U_l, X_l)}}\\
&=
 \prod_{j=0}^J \frac{  \prod_{i \in \mathcal{O}(t^{(j)})} e^{\phi \left(  {U}_i, X_i\right)}}{  \prod_{i \in \mathcal{O}(t^{(j)})} \left\{\sum_{l \in \mathcal{R}(t^{(j)})} e^{\phi \left(  U_l, X_l\right)}- \sum_{m\in \mathcal{O}(t^{(j)}): T_m> T_i} e^{\phi \left( U_m, X_m\right)}
 \right\}}
.
\end{align*}

Having only access to $rec(T_i)$ but not $T_i$, the last  sum in the denominator is not known. We propose an analogue version of the partial likelihood correction for ties presented in \citeA[Section 6, point g]{efron77}: 

$$
\mathcal{L}
= \prod_{j=0}^J \frac{  \prod_{i \in \mathcal{O}(t^{(j)})} e^{\phi \left(  {U}_i, X_i\right)}}{  \prod_{r=0,\dots, O_{j}-1} \left\{\sum_{l \in \mathcal{R}(t^{(j)})} e^{\phi \left(  U_l, X_l \right)}- \frac{r}{{O_j}}\sum_{s\in \mathcal{O}(t^{(j)})} e^{\phi \left( U_s, X_s\right)}
\right\}}.
$$

The heuristic of this approximation is that for every $i\in\mathcal{O}(t^{(j)})$, the number of summands in  $\sum_{m \in \mathcal{O}(t^{(j)}): T_m> T_i} e^{\phi \left( U_m, X_m\right)}$ 
is a unique  $r\in \{0,\dots,O_{j}-1\}$. Hence, the approximation replaces 
each $e^{\phi \left( U_m, X_m\right)}$ in the  summand by  $O_j^{-1}\sum_{s \in \mathcal{O}(t^{(j)})} e^{\phi \left( U_s, X_s\right)}$.

We will work with the negative log-likelihood: 
\begin{align}
  \label{eq:loglkh}
\ell= \sum_{j= 0}^J  \sum_{r=0,\dots, O_{j}-1} \log\left( \sum_{l \in \mathcal{R}(t^{(j)})} e^{\phi \left( {U_l, X_l }\right)}- \frac{r}{O_j}\sum_{s \in \mathcal{O}(t^{(j)})} e^{\phi \left(  U_s, X_s \right)} 
\right) - \sum_{i \in \mathcal{O}(t^{(j)})}\phi \left( {U}_i, {X}_i \right).
\end{align}

\subsubsection{Cox model (COX)}

The proportional model proposed in \citeA{cox72}, uses a linear function to specify $\phi\left(U,X\right)=\theta^TX +\theta_0 U$, with $\theta \in \mathbb{R}^p$ and $\theta_0 \in \mathbb{R}$. To model continuous features that avoid the linear scale, we follow the approach in \citeA{gray92} and introduce splines in the log-risk function \cite{eilers96}. 
We will indicate the coordinates of $X$ with  superscripts. Assuming that within the $p$ features we have $c$ features for the splines and  $p-c$ features for the linear term, the log-risk function is 

$$\phi\left(U,X\right)= \sum^{c}_{s=1} \zeta_s\left(X^s\right) + \theta_0 U+\sum^{p}_{l=c+1} \theta_l X^l ,$$

where, for $V=U$ or $X^l$, $\zeta_v\left(V\right) = \sum_{k=0}^{\kappa_v} \beta^v_k B^v_{k} (V)$ and $\mathbf{\beta}^v=\left(\beta^v_1, \ldots, \beta^v_{\kappa_v}\right) \in \mathbb{R}^{\kappa_v}$. 
Here, $B^v_{k} (V)$ are basis functions and $\kappa_v \in \mathbb{N}$ is the number of knots in the spline. 
The smoothing of $U$ and the other $c$ features that are  modeled with splines are controlled with an additional penalty term in the log-partial likelihood during the model fitting. In the fitting phase we minimize the penalized likelihood 

$$l^p(\theta, \beta^{0}, \dots, \beta^{p})=l + \frac{1}{2} \sum_s \rho_s \int\left[\zeta_s^{\prime \prime}(z)\right]^2 d z, $$

where $\rho_{0}, \ldots, \rho_{p}$ are the parameters controlling the smoothing (no penalty for $\rho=0$ and forcing the spline to a linear form for $\rho=+\infty$). Noting that $\int\left[\zeta_s^{\prime \prime}(z)\right]^2 d z$ is a quadratic form of $\mathbf{\beta}^s$, for some definite positive matrix $\mathbf{P}_s$, the penalized likelihood can be rewritten as 

$$l^p(\theta, \beta^{0}, \dots, \beta^{p})=l +\frac{1}{2} \sum_{s=0}^p \rho_s\beta_s^{\prime} \mathbf{P}_s\beta_s.$$

We minimize $l^p$ for the spline parameters and the $\theta$ parameters. We fit the COX model using the \texttt{R} package \texttt{survival} \cite{R, survivalpackage}, that allows us to specify the smoothing penalty terms and the number of knots in the splines.

\subsubsection{Neural Networks (NN)}

In this model, we extend the approach in \citeA{katzman18} to model left-truncated data while using the correction for ties as introduced earlier. We estimate the log-risk function via a feedforward neural network with a vector of parameters $\theta^{NN} \in \mathcal{P}$. Here,   $\mathcal{P}$ is the space of the possible neural network parameters \cite[p.168]{b:gf16}. 

In the optimization phase, we minimize the regularized objective function

$$
l + \rho \left(\epsilon\|\theta^{NN} \|^2_2+(1-\epsilon)\|\theta^{NN} \|_1\right),
$$

where the hyper parameters $\rho, \epsilon$ allow for the elastic penalty term and the p-norm is $\|\theta^{NN}\|_p=\left(\sum_{\ell}\left|\theta^{NN}_\ell\right|^p\right)^{\frac{1}{p}}$ with $p \in \mathbb{R}, p \geq 1$, and $\ell$ is the index of the parameters of the neural network. The neural network parameters are trained after selecting the hyperparameters through cross-validation. Further details on the neural network hyperparameters can be found in \cref{appendix:bayescv}.

\subsubsection{eXtreme Gradient Boosting (XGB)}

The derivation we present in this section is a modification of \citeA{liu20}, where the authors consider right censoring only and not left-truncation.

The eXtreme gradient boosting algorithm, needs as input the gradient and the second order derivatives of the negative log likelihood function \eqref{eq:loglkh} with respect to ${\phi \left(U_i, X_i\right)}$. The   gradient is
\[g_{i}= e^{\phi \left( U_i, X_i \right)}\left(\upsilon_i -\iota_i\right) -1,
\]

and the second order derivative is
\[h_{i}=g_{i}+1-e^{2 \phi \left( U_i, X_i\right)}\left(\gamma_i-\omega_i\right),\]

where for $rec(T_i)=t^{(l)}$, 

\[
\upsilon_i = \sum^l_{m=0} \sum_{r=0,\dots, O_{m}-1} \frac{1}{ \sum_{k \in \mathcal{R}(t^{(m)})} e^{\phi \left( U_k, X_k \right)}- \frac{r}{O_m}\sum_{s \in \mathcal{O}(t^{(m)})} e^{\phi \left({X}_s, U_s \right)} },
\]

\[
\iota_i = \sum_{r=0,\dots, O_{l}-1} \frac{r/O_l}{\sum_{k \in \mathcal{R}(t^{(l)})} e^{\phi \left({X}_k, U_k ; \theta^{XGB} \right)}- \frac{r}{O_l}\sum_{s \in \mathcal{O}(t^{(l)})} e^{\phi \left( U_s, X_s \right)}},
\]

\[
\gamma_i = \sum^l_{m=0} \sum_{r=0,\dots, O_{m}-1} \frac{1}{ \left(\sum_{k \in \mathcal{R}(t^{(m)})} e^{\phi \left({X}_k, U_k  \right)}- \frac{r}{O_m}\sum_{s \in \mathcal{O}(t^{(m)})} e^{\phi \left( U_s, X_s \right)} \right)^2},
\]

\[
\omega_i = \sum_{r=0,\dots, O_{l}-1} \frac{\left(1-\left(1-\frac{r}{O_l}\right)^2\right)}{ \left(\sum_{k \in \mathcal{R}(t^{(l)})} e^{\phi \left( U_k, X_k \right)}- \frac{r}{O_l}\sum_{s \in \mathcal{O}(t^{(l)})} e^{\phi \left( U_s, X_s \right)}\right)^2}.
\]

As illustrated in \cite{chen16}, XGB is an iterative algorithm, and at each iteration $t$, the current tree further split by optimizing the objective function

$$\sum_{i=1}^n\left[g_{i} f_t\left({X}_i,U_i; \theta_t\right)+\frac{1}{2} h_{i} f_t^2\left({X}_i, U_i;\theta_t\right)\right]+\gamma \theta_t + \frac{1}{2} \rho\|w_t\|^2,$$

where $\gamma>0$ is a penalty term on the tree complexity, and $\rho>0$ is the $\ell_2$ regularization term for the leaf weights $w \in \mathbb{R}^{\tau_t}$.

\subsection{Baseline hazard}
\label{s:baseline}

In this section we discuss the estimation the distribution of the baseline hazard. Following the discussion in \cite{cox72}, once an estimate $\hat \theta$ of $\theta$ is obtained minimising the partial likelihood in \cref{eq:loglkh}, we can derive an estimator for the baseline using the model full-likelihood.

Many implementations of the baseline rely on the  approach in \citeA{breslow74}. There, the author assumes implicitly that the events $[t^{(j)}, t^{(j+1)}) $ occur simultaneously at $t^{(m)}$. In contrast, we will assume that the claims report are uniform distributed within the tie. This makes the estimation of the baseline consistent with the way we will later transform the estimated hazard function into development factors.

Without ties, a histogram estimator for the interval $[t^{(j)}, t^{(j+1)}) $,  $\hat \alpha_{0, t^{(j)}}$, can be obtained by the least-squares minimization criterion
\[
 \underset{\theta}{\textrm{argmin} }\lim_{\varepsilon \downarrow 0} \sum_{i=1}^n
\int_{t^{(j)}}^{t^{(j+1)}}  \left[\left\{   \varepsilon^{-1}\int_{s}^{s+\varepsilon} \mathrm dN_i(u) - \theta e^{\hat \phi(U_i,X_i)}Y_i(s)    \right\}^2
- \kappa(\varepsilon, s)\right] \frac{1}{e^{ \hat \phi(U_i,X_i)}} \mathrm ds,
\]

where $\kappa(\varepsilon, s)= \left(\varepsilon^{-1}\int_{s}^{s+\varepsilon} \mathrm dN_i(u)\right)^2$ is a vertical shift that makes the expression well-defined.
The solution is
\begin{align*}
 \hat \alpha_{0, t^{(j)}}&=    \frac{O_j}{\sum_{i=1}^n \int_{t^{(j)}}^{t^{(j+1)}}  e^{\hat \phi\left( U_i,X_i\right)} Y_i(s) \mathrm ds}\\
&=    \frac{O_j}{\sum_{i=1}^n \int_{t^{(j)}}^{t^{(j+1)}}  e^{\hat \phi\left( U_i,X_i\right)} I(T_i \leq s < \mathcal U-U_i) \mathrm ds} \\
&=    \delta^{-1}\frac{O_j}{\sum_{l \in \mathcal{R}(t^{(j)})} e^{\hat \phi\left(U_l, X_l\right)} - \eta \sum_{m \in \mathcal{O}(t^{(j)})} e^{\hat \phi\left(U_m,X_m\right)}
},
\end{align*}

with unknown constant $\eta$ $\in [0,1]$, because only $rec(T_i)$ is observed and not $T_i$.

As default, we propose $\eta=0.5$ as estimator:
\begin{align}
\label{est}
    \hat \alpha_{0, t^{(j)}}=\delta^{-1}\frac{O_j}{\sum_{l \in \mathcal{R}(t^{(j)})} e^{\phi\left(U_l, X_l\right)} - \frac 12 \sum_{m \in \mathcal{O}(t^{(j)})} e^{\phi\left(U_m,X_m\right)}
    },
\end{align}
corresponding to the expectation of the above expression under assumption that claim reports occur uniformly on the interval
$[t^{(m)}, t^{(m+1)})$.

Similar to the assumption that $U_i$ is discrete, this assumption is rather innocent if $\delta=t^{(m)}- t^{(m+1)}$ is small,
e.g. one day, but  can be problematic if the intervals are larger.

\section{\texorpdfstring{Modeling within development triangles}{Modeling within development triangles}}
\label{s:modelingtheclaimsdev}

In this section, we explain the connection between continuous individual hazard rates and chain-ladder development factors.  We start the section with the definition of development triangles. We then connect the continuous time framework with the observation of ties that we used for model estimation to the discrete setting that we will construct for model predictions. A byproduct of our proposal is that for different level of granularity, say yearly or quarterly, the predictions do not change. In the sequel we will sometimes write  $\hat \alpha(j|k,x)$ for the estimator $\hat \alpha(t^{(j)}|u^{(k)},x)$ derived in the previous section.

\subsection{From hazard rates to development factors}
We start by dividing the square  $[0, \mathcal U] \times 
[0,\mathcal T]$ into small, feature dependent parallelograms.
For $k=0,\dots, K-1; j=0,\dots, J-1$ and $x \in \mathbb{R}^p$, we  define 

\[
\mathcal{P}_{k,j}(x)= \{(t,u,x): t^{(j)}+u^{(k)}-u\leq t \leq t^{(j+1)}+u^{(k)}-u; u\in [u^{(k)},u^{(k+1 )}), t\geq 0\},
\]
where $u^{(k)}=k\delta$. The individual reported claims are grouped as

\[
O_{k,j}\left(x\right)=\sum_i \int I\left(\left( s,U_i, X_i\right) \in P_{k,j}\left(x\right)\right) d N_i(s).
\]
Note that $O_{k,j}(x)$ is fully observed on the upper triangle,
$k+j \leq K-1$, i.e., it is not impacted by the conditioning of
$T_i+U_i<\mathcal U$.
Furthermore $O_{k,j}(x)$ can be calculated with
knowledge only about $rec(T_i)$ rather than $T_i$.
Lastly, $O_{k,j}(x)$ is zero in the lower triangle,
$k+j >K-1$ because $T_i+U_i<\mathcal U$.

In reserving, the raw age-to-age factors $\tilde {f}_{k,j}(x)=\sum_{\ell \leq j} O_{k \ell}\left(x\right)/\sum_{\ell < j} O_{k \ell}\left(x\right)$ are well-known objects, but they cannot be used for prediction because they are not defined on the un-observed lower triangle $k+j>K$. Even ignoring this problem, $\tilde {f}_{k,j}(x)$ is too noisy in describing the development from $j-1$ to $j$ because it is calculated separately for every $k$ and $x$.
We propose to use the hazard rate estimated in the previous section to derive a more stable estimate of the development from $j-1$ to $j$.
For $k=1,\dots, K-1; j=1,\dots, J-1$, we propose to estimate the development factors as
\begin{equation}
\label{eq:dftohr}
\hat {f}_{k,j}(x) := \frac{2+ \delta \hat \alpha(j|k,x) }{2- \delta \hat \alpha(j|k,x)}.
\end{equation}
The formula can be motivated from the fact that equation \eqref{eq:dftohr} is true
if $\hat {f}_{k,j}(x)$ is replaced by the raw age-to-age $\tilde {f}_{k,j}(x)$ and
$\delta\hat \alpha(j|u,x)$ is replaced by the raw observations of the ratio of occurrence and exposure $(O_{k,j}(x)/\left(\sum_{l<j}O_{k,l}(x)- \frac 1 2 O_{k,j}(x)\right))$:

\[
\tilde {f}_{k,j}(x)= \frac{2+ O_{k,j}(x)/\left(\sum_{l<j}O_{k,l}(x)- \frac 1 2 O_{k,j}(x)\right)}{2- O_{k,j}(x)/\left(\sum_{l<j}O_{k,l}(x)- \frac 1 2 O_{k,j}(x)\right)}.
\]

In other words, equation \eqref{eq:dftohr} represents a plug-in estimator. One problem that may arise  with equation \eqref{eq:dftohr} is that the estimator blows up if $\hat \alpha(j|k,x)$ is close to $2\delta^{-1}$. This can happen if $\delta$ is not chosen small enough. If  $\delta$ can not be chosen smaller due to data quality issues, one can try to modify equation
 \eqref{est} by replacing the 1/2 factor with a different estimator for $\eta$, $\hat \eta$ ,thereby changing the uniform assumption
 to a possibly more suitable one. In that case, equation \eqref{eq:dftohr} becomes
 \[
\hat f_{k,j} :=  \frac{1 +(1-\hat \eta)\delta \hat \alpha(j|k,x)}{1-\hat \eta  \delta \hat \alpha(j|k,x)},
 \]
due to the identity

\[
\tilde {f}_{k,j}(x)= \frac{\sum_{l\leq  j}O_{k,l}(x)}{\sum_{l<j}O_{k,l}(x)}=\frac{1+(1-\eta) \; O_{k,j}(x)/\left(\sum_{l<j}O_{k,l}(x)- \eta \; O_{k,j}(x)\right)}{1-\eta \; O_{k,j}(x)/\left(\sum_{l<j}O_{k,l}(x)- \eta \; O_{k,j}(x)\right)},
\]
for any $\eta \in [0,1]$.
 However, we caution that changing $\hat \eta$ to a more ``favorable'' value may just hide the bias due to a too large $\delta$. If enough data is available performance of the estimator can be checked 
 by back-testing the obtained predictions.
 In our application $\delta$ equaling one day was small enough and no modification was needed.

\subsection{Predictions into the lower triangle}
\label{ss:predictions}

We define the cumulative entry at development time $j=0,\dots, J-1$ for accident period $k=0,\dots, K-1$ with features $x$ as

$$C_{k,j}\left(x\right)=\sum_{l\leq j}O_{k,l}(x).$$

In the lower triangle, i.e.,  $j+k>K-1$, our estimator for the number of reports at development time $j$ for accident period $k$ is

\begin{align*}
\hat{O}_{k,j}(x) = \begin{cases}
    {C}_{k,K-k-1}(x)  \left( \hat{f}_{k,l}(x) - 1 \right) & \text{if}\  j=K-k,\\
    {C}_{k,K-k-1}(x)  \left( \prod_{l=K-k}^{j} \hat{f}_{k,l}(x) - \prod_{l=K-k}^{j
    -1} \hat{f}_{k,l}(x) \right) & \text{if}\   j > K-k.
    \end{cases}
\end{align*}

\subsection{Increasing the granularity of development factors}

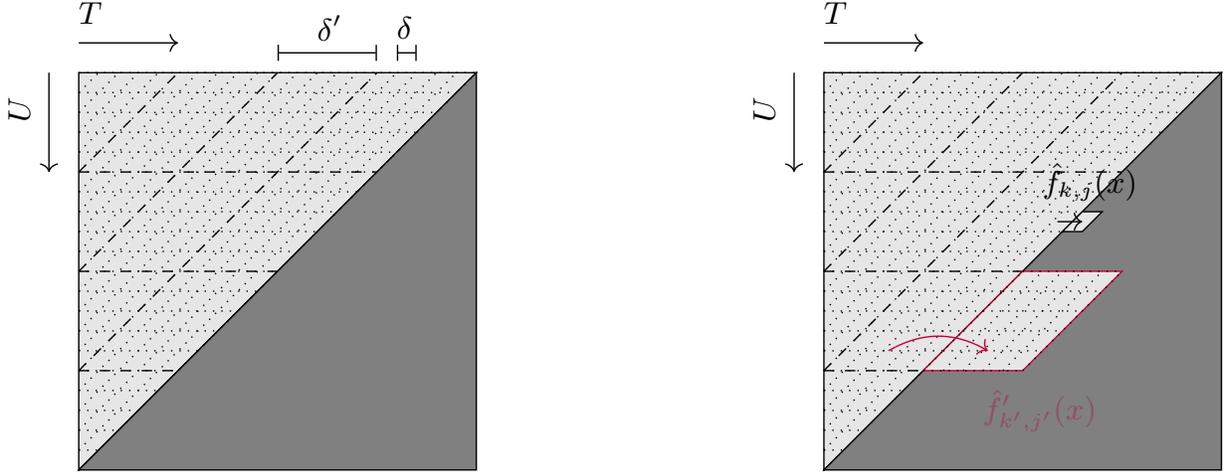
\begin{figure}[ht]
\small\centering
\tikzset{cross/.style={cross out, draw=black, minimum size=2*(#1-\pgflinewidth), inner sep=0pt, outer sep=0pt},
cross/.default={1pt}}

\begin{subfigure}[t]{0.4\textwidth}
\centering
\resizebox{\linewidth}{!}{
\begin{tikzpicture}

\coordinate (r0) at (0,4);
\coordinate (s0) at (4,4);
\coordinate (si) at (4,0);

\coordinate (s0l) at (0,0);
%

\filldraw[draw=black, fill=gray!20] (s0l) -- (r0) -- (s0) -- cycle;

\filldraw[draw=black, fill=gray] (s0) -- (si) -- (s0l) -- cycle;

\foreach \x in {1,2,3}{
\draw [dashed] (0,\x) -- (\x,\x);
}

\foreach \x in {1,2,3}{
 \draw [dashed] (0,\x) -- (4-\x,4);
}

\foreach \x in {0,0.2,0.4, ..., 4}{
\draw [dotted] (0,\x) -- (\x,\x);
}

\foreach \x in {0,0.2,0.4, ..., 4}{
 \draw [dotted] (0,\x) -- (4-\x,4);
}



\draw[->] (-.3,4) to (-.3,3);
\node[text width=1cm,rotate= 90] at (-.6,4) 
    {$U$};

\draw[->] (0,4.3) to (1,4.3);
\node[text width=4cm] at (2,4.6) 
    {$T$};

\draw[|-|](3.2,4.2) to (3.4,4.2);
\node[text width=.1cm] at (3.25,4.45) 
    {$\delta$};

\draw[|-|](2,4.2) to (3,4.2);
\node[text width=1cm] at (2.9,4.45) 
    {$\delta^\prime$};

\end{tikzpicture}}
\subcaption{\label{fig:twogranularities} development triangle showing two levels of granularity ($\delta$ and $\delta^\prime$). The data in the parallelograms with granularity $\delta$ are included in the parallelograms with granularity $\delta^\prime$, with $\delta^\prime>\delta$ and $\lfloor \frac{\delta^\prime}{\delta}\rfloor =0$. \newline}
\end{subfigure}
\hfill
\begin{subfigure}[t]{0.4\textwidth}
\resizebox{\linewidth}{!}{
\begin{tikzpicture}

\filldraw[draw=black, fill=gray!20] (s0l) -- (r0) -- (s0) -- cycle;

\filldraw[draw=black, fill=gray] (s0) -- (si) -- (s0l) -- cycle;

\foreach \x in {1,2,3}{
\draw [dashed] (0,\x) -- (\x,\x);
}

\foreach \x in {1,2,3}{
 \draw [dashed] (0,\x) -- (4-\x,4);
}

\foreach \x in {0,0.2,0.4, ..., 4}{
\draw [dotted] (0,\x) -- (\x,\x);
}

\foreach \x in {0,0.2,0.4, ..., 4}{
 \draw [dotted] (0,\x) -- (4-\x,4);
}

\draw[->] (-.3,4) to (-.3,3);
\node[text width=1cm,rotate= 90] at (-.6,4) 
    {$U$};

\draw[->] (0,4.3) to (1,4.3);
\node[text width=4cm] at (2,4.6) 
    {$T$};

\coordinate (r0) at (1,1);
\coordinate (s0) at (2,1);
\coordinate (si) at (3,2);

\coordinate (s0l) at (2,2);
%
\filldraw[draw={rgb:red,78;green,0;blue,22}, fill=gray!20] (r0) -- (s0) -- (si)-- (s0l) -- cycle;

\filldraw[draw=black, fill=gray!20] (2.4,2.4) -- (2.6,2.4) -- (2.8,2.6) -- (2.6,2.6) -- cycle;

\foreach \x in {1,1.2,1.4, ..., 2}{
\draw [dotted] (\x,\x) -- (\x+1,\x);
}

\foreach \x in {1,1.2,1.4, ..., 2}{
 \draw [dotted] (\x,1) -- (\x+1,2);
}

\draw[->] (2.35,2.5) to (2.6,2.5);
\draw[->, color={rgb:red,78;green,0;blue,22}] (.65,1.2) to [bend left] (1.65,1.2);

\node[text width=1cm] at (2.7,2.9) 
    {$\hat f_{k,j}(x)$};
\node[text width=1cm] at (2.1,0.6) 
    {\textcolor{myred}{$\hat f^\prime_{k^\prime,j^\prime}(x)$}};

\end{tikzpicture}}

\subcaption{\label{fig:dfs} Using the correspondence between development factors and average hazard rates we can predict the IBNR development on granularity $\delta$ (black). Afterwards, we can produce development factors for granularity $\delta^\prime$ (red). }

\end{subfigure}

\caption{\label{fig:granularities} Our approach can handle different levels of granularities (left hand side). Starting from the individual data we can easily produce results for different aggregation levels (right hand side).}
\end{figure}

In the previous sections, we defined the development factors as discrete objects that we use for projecting the future reports given some granularity $\delta=t^{(j)}-t^{(j-1)}=u^{(k)}-u^{(k-1)}$. In this section, we want to elaborate on the possibility of choosing different values of $\delta^\prime > \delta$ for the computation of the development factors.
 
For example in the first data application of this paper we will start from daily data but we will be interested in reporting the results in quarterly and yearly flows. In \cref{fig:twogranularities} we provide an intuition behind the change of granularity. For simplicity, let us assume that $\delta^\prime/\delta, (K-1)(\delta/\delta^\prime), (J-1)(\delta/\delta^\prime)$, are  
integer valued. 

For $k=0,\dots, K-1; j=0,\dots, J-1$, the original  parallelograms $P_{k,j}(x)$ are with respect to some granularity $\delta$. 

Now, for $k=0,\dots, (K-1)(\delta/\delta^\prime)-1=K^\prime-1, j=0,\dots, (J-1)(\delta/\delta^\prime)-1=J^\prime-1$, we  define the parallelogram and fitted occurrences on the granularity $\delta^\prime$ as:
\begin{align*}
\mathcal I(k,j)&= \left\{ (h,l): h \in \Big\{ k\cdot \frac{\delta^\prime}{\delta},\ldots,(k+1)
\cdot \frac{\delta^\prime}{\delta} -1 \Big\} ,\, l \in \Big\{ j\cdot \frac{\delta^\prime}{\delta},\ldots,(j+1)
\cdot \frac{\delta^\prime}{\delta} -1 \Big\} \right\},\\
P'_{k,j} (x) &= \bigcup_{(h,l)\in \mathcal I(k,j)} P_{h,l},\\
\hat O'_{k,j} (x) &= \sum_{h,l \in \mathcal I(k,j)} \hat O_{h,l}, \quad k+j > K^\prime-1.
\end{align*}

The occurrence for granularity $\delta^\prime$ is $O^\prime_{k,j}\left(x\right)=\sum_i I\left(\left( T_i, U_i, X_i\right) \in P^\prime_{k,j}\left(x\right)\right)
$ and the corresponding cumulative entry for  granularity  $\delta^\prime$ becomes $C^\prime_{k,j}\left(x\right)=\sum_{s\leq j}O^\prime_{k,s}(x)$. 
To derive the development factors for granularity
$\delta^\prime$, we first need to define the following intermediate quantities:
\begin{align*}
\tilde {O}_{k,j}(x) &=
 {O}_{k,0}(x) \prod_{l={1}}^{j} \hat{f}_{k,l}(x), \\
\tilde O^\prime_{k,j} \left( x\right) &= \sum_{(h,l)\in \mathcal I(k,j)}\tilde O_{h,l}.
\end{align*}

We can use these quantities to obtain development factors with granularity $\delta^\prime$:
\begin{align*}
\hat{f}^\prime_{k,j}(x) = 
   \frac{ \sum_{l \leq j } \tilde{O}^\prime_{k,l}(x)}{  \sum_{l < j }  \tilde O^\prime_{k,l}(x) }.
\end{align*}

In \cref{fig:dfs} we illustrate the idea behind using the development factors $\hat{f}^\prime_{k,j}(x)$ on the granularity $\delta^\prime$ to forecast future reports.

\section{Models comparison}
\label{sec:evaluation}

In this Section, we will first define three performance metrics to compare our models (COX, NN, and XGB) to the chain-ladder \cite<CL, >{mack93} and our implementation of the methodology presented in  \cite<MW, >{wuthrich18}. Secondly, in \cref{ss:crps} we will describe the Continuously Ranked Probability Score \cite{gneiting07}, which will be used to rank the individual models. 

In this phase, we use the development factors that we obtained to estimate the total future notifications $\sum_x O^\prime_{k,j}(x)$, for ${j,k: k+j>K-1}$ for $k=0,\dots, K-1; j=0,\dots, J-1$. We first compare in absolute and relative terms the total predicted reportings to the actual reportings, using the $ R^{\text{TOT}}$ metric for the total error that we define as

\begin{align}
\label{eq:eir}
    R^{\text{TOT}} = |1 - \frac{\sum_{j,k: k+j>K-1}  \sum_x \hat O^\prime_{k,j}(x) }{\sum_{j,k: k+j>K-1} \sum_x  O^\prime_{k,j}(x)}|.
\end{align}

Secondly, we compare the the predicted reportings to the actual reportings cell-wise, using $R^{\text{cell-wise}}$ metric. For some granularity $\delta^\prime$, we calculate

\begin{equation}
\label{eq:retot}
R^{\text{cell-wise}} =  \frac{\sum_{j,k:k+j>K-1} |\sum_x O^\prime_{k,j}(x)- \sum_x \hat O^\prime_{k,j}(x)|}{\sum_{j,k:k+j>K-1} \sum_x  O^\prime_{k,j}(x)}.
\end{equation}

Lastly, given some granularity $\delta^\prime$, we perform the same evaluation for each of the calendar times $\tau=K', \ldots, 2K'-2$, using the $R^{\text{cal-wise}}$ metric,

\begin{equation}
\label{eq:recal}
R^{\text{cal-wise}}=\frac{\sum^{2K-2}_{\tau=K} | \sum_{j,k:k+j=\tau } \left( \sum_x  O^\prime_{k,j}(x)- \sum_x \hat  O^\prime_{k,j-1}(x)\right)|}{\sum_{j,k:k+j>K-1} \sum_x  O^\prime_{k,j}(x)}.
\end{equation}

\subsection{Continuously Ranked Probability Score}
\label{ss:crps}

The $R^{\text{cell-wise}}$ and the $R^{\text{cal-wise}}$ are metrics that compare our models in terms of predicted counts to the realized counts. However, they do only take point estimates into consideration and not the predicted distribution. In this section, we consider as additional metric a scoring rule that takes the predicted distribution into account \cite{gneiting07}. When a set of different stochastic models is available, scoring rules can be used as a criterion to assesses the quality of the forecasts by assigning to each model a score. By giving better scores to models that provide better forecasts, we can rank competing forecast procedures.

A scoring rule is a function $s(f,y)$ taking values in the real line $\left[0, \infty \right)$ where $f$ is a density forecast and $y \in \mathbb{R}$ is a future realization from the conditional sampling distribution $Y$. In this Section, we propose the Continuously Ranked Probability Score (CRPS).

For some individual $i$ belonging to the out-of-sample data (lower triangle) $i= n+1, \ldots, n+m$, with $\text{rec}(T_i) = t^{(j)}$, we define the individual CRPS as: 

$$
\text{CRPS}_{i} = \sum^{j-1}_{z=1} (1-\hat{S}_{k,z}(x_i))^2 \Delta_z  + \sum^{J}_{z=j+1}  \hat{S}^2_{k,z}(x_i) \Delta_z + \frac{1}{2}\Delta_j(\hat{S}^2_{k,j}(x_i)+(1-\hat{S}_{k,j}(x_i))^2) ,
$$

with $\Delta_z = t^{(z)} - t^{(z-1)}$ and the predicted survival function 
\begin{align}
\label{eq:estimatedsf}
\hat{S}_{k,j}(x_i)=\frac{1}{\prod^{j}_{l=1} \hat{f}_{k,l}(x_i)}.
\end{align}

As a performance metric, we adopt the average individual CRPS in the out-of-sample data

$$
\text{CRPS} = m^{-1} \sum^{n+m}_{i^\prime=n+1} \text{CRPS}_{i}.
$$
The CRPS is taken to be negatively oriented, meaning that the lowest score indicates the better model. 

Furhter diagnostics on the individual model fits can be performed by inspecting the partial log-likelihood. We will report the partial log-likelihood of our models for the different data sets that we inspect in this manuscript in \cref{appendix:llminimization}. The model with the minimum in-sample average negative partial log-likelihood is the model that fits best the training data. In this context, the log-likelihood can only provide a reference value to understand what model is capable to minimize the objective function on the in-sample data during the fitting, i.e. \Cref{eq:loglkh}.

We remark that both the CL and MW models are based on discrete formulations of the development factors. Within their framework, computing a survival function is not meaningful. Henceforth, we will not compute the CRPS for MW and CL in the application section of this paper.

\section{Data Application on simulated data}
\label{sec:dataapplicationsim}

In this section, we assess our modeling approach to six distinct simulated scenarios. The composition of the data in each scenario is detailed in \cref{tab:simulation_composition}. For each scenario, we generate 20 independent data sets. Model performance is evaluated for each replication using the scoring rules outlined in \cref{sec:evaluation}. The results reported are averages across the 20 replications. We consider this number of repetitions sufficient to reduce the influence of stochastic variability in the simulation outcomes. According to the experience of the authors in the actuarial practice, we generated scenarios that could occur in the real world. Additional details on the simulation algorithm are in \cref{appendix:simulation}.

\begin{table}[ht]
\centering
\small
\begin{tabular}{p{2.5cm}|l|l|l}
Scenario & Features & Description & Data Size (expected) \\ \hline
\multirow{4}{2cm}{Alpha, Beta, Gamma, Delta, Epsilon} 
& \texttt{claim\_number} & Policy identifier. & 
\multirow{4}{3.2cm}{28800  (Alpha, Gamma, Delta, Epsilon)\\23620 (Beta)}
 \\
& \texttt{claim\_type} $\in \left\{0, 1 \right\}$ & Type of claim, categorical. & \\
& \texttt{AD} $\in \{1, \ldots, 1440\}$ & Day of accident, integer valued. & \\
& \texttt{DD} $\in \{1, \ldots, 1440\}$& Delay (in days) from accident to reporting, integer valued. & \\ \hline

\multirow{6}{2cm}{Zeta}
& \texttt{claim\_number} & Policy identifier. & 
\multirow{6}{3.2cm}{9448} \\
& \texttt{age} $\in \{50, \ldots, 55\}$ & Age of the policy holder. Integer valued. & \\
& \texttt{property\_value} $\in \mathbb{R}^+$ & Value of the insured property. Continuous. & \\
& \texttt{business\_use} $\in \{\text{Y}, \text{N}\}$ & Whether the property is for business use. Categorical. & \\
& \texttt{AD} $\in \{1, \ldots, 1440\}$ & Day of accident, integer valued. & \\
& \texttt{DD} $\in \{1, \ldots, 1440\}$& Delay (in days) from accident to reporting, integer valued. & \\ \hline
\end{tabular}
\caption{\label{tab:simulation_composition} We illustrate the data composition (column two) in each scenario (column one). The features are described in column three. In column four we write the (expected) number of observations simulated in each data set for the different scenarios: The number of claims is randomly generated in the different accident periods.}
\end{table}

We compare our models (COX, NN, and XGB) with the chain-ladder \cite<CL, >{mack93} and the neural-networks based approach in \cite<MW, >{wuthrich18}. 
From our experience in the industry and in particular on the data from \cref{sec:dataapplicationrd}, when implementing the CL, actuaries assume homogeneous data and aggregate together the different types of claims in one triangle, disregarding the covariate information. We follow this approach in the paper. Further details on the MW approach and how we implemented it can be found in Appendix \ref{appendix:mw}.

\subsection{Six simulated scenarios}

This section provides an overview of the simulations. The data are simulated on a daily grid on a $4$ years time horizon, meaning that we will observe up to $1440$ accident days ($\mathcal{T}=1440$). We provide results on a quarterly grid and compare our models to the CL and the neural-networks based approach from MW. 

The first five scenarios have some common characteristics: 

\begin{itemize}
    \item A mix of \texttt{claim\_type 0} and  \texttt{claim\_type 1}. The parameters are chosen to make \texttt{claim\_type 0} resemble property damage and \texttt{claim\_type 1} bodily injuries.
    \item Bodily injuries (\texttt{claim\_type 1}) are longer tailed, meaning that their resolution takes more time than property damage (\texttt{claim\_type 0}), see \cite{ajne94}.
\end{itemize}

We name the $5$ scenarios Alpha, Beta, Gamma, Delta, Epsilon. 

The simulations were performed using the \texttt{R} package \texttt{SynthETIC}, see \cite{Avanzi21}. For each claim type, data is simulated in a continuous setting. To imitate a real world portfolio, we populate the data with an accident day (\texttt{AD}) and development day (\texttt{DD}) for each observation.
In the spirit of \cite{Avanzi21}, for each claim type, we model occurrence and development of claims in two modules:

\begin{itemize}
    \item We first fix the total amount of individuals at risk in the portfolio for each accident day. We then let the actual frequency of claims being reported in the accident date be randomly simulated (we draw it from a Poisson distributed random variable with rate $0.05$). In scenarios Alpha, Gamma, Delta and Epsilon the individual at risk per \texttt{AD} are in the same proportion for each \texttt{claim\_type}, i.e. in each \texttt{claim\_type} we have $200$ individuals at risk per \texttt{AD}. In scenario Beta the individuals at risk are decreasing by one unit every $10$ \texttt{AD}, e.g. we will have $200$ individuals at risk in \texttt{AD}$= 1$, $199$ individuals at risk in \texttt{AD}$= 11$ and only $56$ individuals at risk in \texttt{AD}$=1440$. The total (expected) data size, comprehensive of train set and test set is $28800$ for scenarios  Alpha, Gamma, Delta and Epsilon and $23620$ for scenario Beta.

    \item We then simulate the reporting delay of the individual claims. As we need a proportional hazard structure in reverse development time, we need to be careful when simulating the reporting delays. The proportional hazard is
   
    $$
    \alpha(t|U=\text{\texttt{AD}}, {X}=\text{\texttt{claim\_type}})= \alpha_0(t)e^{\phi\left(\texttt{AD},\texttt{claim\_type}\right)}.
    $$

The parametrization of $\alpha_0(t)$ and $\phi\left(\texttt{claim\_type},\texttt{AD}\right)$ changes in the different scenarios. In this section we will provide qualitative details on how we designed the different scenarios. We report the parameters we used for the simulation, as well as the scenario-specific functional form of $\phi\left(\texttt{claim\_type},\texttt{AD}\right)$ in \autoref{appendix:simulation}.
    
\end{itemize}

To account for the proportional structure of the hazard, we simulate from a Right Truncated Fréchet-Weibull distribution (RTFWD) \cite{teamah19}. The RTFWD has a five parameter structure $\nu, \pi, \xi, k, b$, and is defined with $0<t \leq b$ with distribution function

$$
F\left(t\right) = e^{-\pi^{\nu}\xi^{\nu k}(t^{-\nu k}-b^{-\nu k})},
$$
With the reverse time hazard 
$$
\alpha(t) = \frac{f(t)}{F(t)},
$$
we get, in the RTFWD case, 
$$
\alpha(t) = \frac{\nu k \pi^{\nu}\xi^{\nu k}(t^{-1-\nu k}) e^{-\pi^{\nu}\xi^{\nu k}(t^{-\nu k}-b^{-\nu k})}}{e^{-\pi^{\nu}\xi^{\nu k}(t^{-\nu k}-b^{-\nu k})}} = \nu k \pi^{\nu}\xi^{\nu k}(t^{-1-\nu k}).
$$

Let us set 

$$\xi=\xi(\text{\texttt{AD}},\text{\texttt{claim\_type}})=\xi_0 (e^{\phi\left(\text{\texttt{AD}},\text{\texttt{claim\_type}}\right)})^{\frac{1}{\nu k}},$$ 

the reverse hazard structure now becomes

\begin{align}
\label{eq:RTFWD}
\alpha(t|\text{\texttt{AD}},\text{\texttt{claim\_type}}) = \alpha_0(t)e^{\phi\left(\text{\texttt{AD}},\text{\texttt{claim\_type}}\right)},
\end{align}

with baseline hazard $\alpha_0(t)=\nu k \pi^{\nu}\xi_0^{\nu k}(t^{-1-\nu k})$;

All simulations are truncated at  $b=1440$, resulting in $4$ years worth of data. 
The parameters of each scenario we set up for the simulation are specified in \autoref{tab:rtfwd1}. The scenarios Alpha, Beta, Gamma, and Delta have the same proportional hazard structure given by \eqref{eq:RTFWD}
with $\text{\texttt{claim\_type}}\in \left\{0,1\right\}$ and $\text{\texttt{AD}}\in \left\{1,\ldots, 1440\right\}$.

The scenarios have the following distinctive traits:

\begin{itemize}
    \item \textbf{Scenario Alpha}: this scenario is a mix of \texttt{claim\_type 0} and  \texttt{claim\_type 1} with same number of claims at each accident period (i.e. the claims volume). As we have an effect based on the claim type and the reporting of the claims only depends on the baseline, in this scenario the CL assumptions are satisfied.     
    
    \item \textbf{Scenario Beta}: we simulate using the same proportional risk component $y$ as \textbf{scenario Alpha}, but the volume of \texttt{claim\_type 1} is decreasing in the most recent \texttt{AD}. When the longer tailed bodily injuries have a decreasing claim volume, aggregated CL methods will overestimate reserves, see \cite{ajne94}. 
    
    \item \textbf{Scenario Gamma}: an interaction between \texttt{claim\_type 1} and accident period (\texttt{AD}) affects the reporting pattern. One could imagine a scenario, where a change in consumer behaviour or company policies resulted in different reporting patterns over time.  For the last simulated accident day, the two reporting delay distributions will be identical. The interaction makes the COX model assumption not valid. In addition, the accident period effect makes the CL model assumptions not valid.
    
    \item \textbf{Scenario Delta}: a seasonality effect dependent on the accident days for \texttt{claim\_type 0} and \texttt{claim\_type 1} is present. This could occur in a real world setting with increased work load during winter for certain claim types, or a decreased workforce during the summer holidays. The presence of an accident period effect makes the CL assumptions not satisfied.
    
    \item \textbf{Scenario Epsilon}: the data generating process violates the proportional hazard assumption. We will generate the data assuming that a) there is an effect of the features on the baseline and b) the proportionality assumption is not valid. Conversely, the CL model is satisfied. In scenario Epsilon, the hazard we simulate from is

    $$
    \alpha(t|\text{\texttt{AD}},\text{\texttt{claim\_type}})= \alpha_0(t|\text{\texttt{AD}},\text{\texttt{claim\_type}})(e^{\phi\left(\text{\texttt{AD}},\text{\texttt{claim\_type}}\right)}+\frac{\phi\left(\text{\texttt{AD}},\text{\texttt{claim\_type}}\right)}{2}),
    $$

    with $\alpha_0(t|x)=0.5 \sqrt{0.1\left(0.1(2+\frac{\phi\left(\text{\texttt{AD}},\text{\texttt{claim\_type}}\right)}{2})t^{-1}\right)}$. 
    
\end{itemize}

The relevant features on the proportional risk part of the hazard are reported in column two of \cref{tab:effectsforscenarios}. In columns four to seven, we indicate with a check mark whether the assumptions for models in the data application (CL, COX, NN, XGB, and MW) are satisfied.

\begin{table}[ht]
\centering
\begin{tabular}{l|l|l|l|l|l|l}
 Scenario & Effect(s) on $\phi$ & CL& COX& NN& XGB& MW\\ \hline
Alpha & \texttt{claim\_type} &\cmark &\cmark&\cmark&\cmark& $\text{\cmark}^*$\\
Beta & \texttt{claim\_type}&\xmark &\cmark&\cmark&\cmark& $\text{\cmark}^*$\\
Gamma & \texttt{claim\_type} + $I\left(\texttt{claim\_type}=1\right)$:$\sqrt{\texttt{AD}}$&\xmark &\xmark&\cmark&\cmark&\xmark \\
Delta & \texttt{claim\_type} +\texttt{AD}&\xmark &\cmark&\cmark&\cmark& \xmark\\
Epsilon & \texttt{claim\_type}&\cmark &\xmark&\xmark&\xmark& $\text{\cmark}^*$\\
Zeta & --- &\xmark &\xmark&\xmark&\xmark& $\text{\cmark}^*$\\
\hline
  \end{tabular}
\caption{\label{tab:effectsforscenarios} The relevant features affecting the proportional risk component (column two) in the scenarios Alpha, Beta, Gamma, Delta, Epsilon and Zeta (column one). For scenario Zeta, column two is left blank since the data are generated using a different model assumption compared to the first five scenarios, see the Transformed Gamma distribution in \cref{eq:zetassumption}}. In order to simplify the reading of this manuscript, we use the notation used in \protect\cite{R}. The effect terms are added with the operator $+$ (plus), the interaction terms are added with the operator $:$ (columns). In columns four to seven, we use a check mark if the models assumptions are satisfied in the scenario. The star around the check marks for the MW model indicates that the assumptions are satisfied as long as $C_{k,j}(x)>0$ for $\{j,k:j+k=K\}$ with $j=0,\ldots, J$ and $k=0,\ldots, K$.
\end{table}

Lastly, we consider  \textbf{Scenario Zeta}, simulated using the \texttt{Synthetic} package for mimicking a rental property insurance portfolio. The delay from accident to report $t \in \mathbb{R}^+$ follows a Transformed Gamma distribution with parameters $s_1,s_2,o >0$ and density

\begin{equation}
\label{eq:zetassumption}
f(t|s_1,s_2,o)=\frac{s_2 (t/o)^{s_1s_2} e^{-(t/o)^{s_2}}}{t \Gamma(s_1)}.
\end{equation}

The covariates included in scenario Zeta are the age of the landlord underwriting the insurance policy (\texttt{age} $\in \{50, \ldots, 55\}$), the value of the insured property (\texttt{property\_value} $\in \mathbb{R}^+$), and an indicator of whether the property is for business use (\texttt{business\_use} $\in \{\text{Y}, \text{N}\}$). In scenario \textbf{Scenario Zeta}, the parameters $s_1$ and $s_2$ will depend on \texttt{property\_value} and \texttt{business\_use} as described in \cref{appendix:simulation}. The age of the landlord is generated to be non-informative, meaning that it will not have an effect on the distribution of the reporting delay. The value of the property insured (\texttt{property\_value})  was generated from a lognormal distribution calibrated on the house prices of the Boston Housing Prices data set as available from the \texttt{R} package \texttt{A3}. Similarly to scenario Beta, in scenario Zeta the individuals at risk are decreasing by one unit every accident date and the rate of the Poisson distribution generating the number of claims every accident date is $0.02$.

\subsection{Results: forecasting using the development factors}

In this section we compare the chain ladder with our models in the five simulated scenarios. In each scenario, we simulate $20$ data sets. Below we describe in brief the procedure that we used.

\begin{enumerate}
    \item For each scenario (Alpha, Beta, Gamma, Delta, Epsilon, and Zeta), we simulate $20$ data set.
    \item The data is pre-processed, see \cref{appendix:datapp}. The categorical covariates are dummy encoded using the \texttt{R} package \texttt{fastDummies} \cite{fastdummies}. The continuous covariates are scaled using a minimum-maximum transformation, similarly to \cite{wuthrich18}.
    \item On each data set, we find the optimal hyper parameters for NN, XGB, and MW using Bayesian optimization \cite{snoek12}. More details on the optimization algorithm can be found in \cref{appendix:bayescv}. 
    \item On each data set, we fit COX and the optimal NN, XGB, and MW. The MW model is fit both on data aggregated monthly and quarterly.
    \item We find the estimated development factors and expected number of future occurrences according to each model. 
    \item We evaluate the performances for COX, NN, XGB, and MW using the $R^{\text{cell-wise}}$,  the CRPS and the $R^{\text{cal-wise}}$ on quarterly basis. The $R^{\text{cell-wise}}$ and $R^{\text{cal-wise}}$ are also computed for the CL.
    \item For each scenario we report the average and the standard deviation of the performance measures of CL, COX, NN, XGB, and MW over the 20 data sets.
    
\end{enumerate}

\begin{table}[ht]
\begin{tabular}{l l c c c c}
\toprule
Model & Scenario & $\texttt{EI}^{\texttt{R}}$ & $R^{\text{cell-wise}}$ & $R^{\text{cal-wise}}$ & CRPS (average) \\

\midrule
CL (\cmark)     & \multirow{6}{*}{Alpha}   & $\textbf{0.0032}\ (\pm\ 0.0250)$ & $\textbf{0.1315}\ (\pm\ 0.0157)$ & $\textbf{0.0452}\ (\pm\ 0.0118)$ & -- \\
COX (\cmark)    &                           & $0.0134\ (\pm\ 0.0455)$ & $0.1373\ (\pm\ 0.0131)$ & $0.0849\ (\pm\ 0.0460)$ & $374.38\ (\pm\ 5.94)$ \\
NN (\cmark)     &                           & $0.0212\ (\pm\ 0.0590)$ & $0.1400\ (\pm\ 0.0216)$ & $0.0976\ (\pm\ 0.0484)$ & $377.27\ (\pm\ 8.91)$ \\
XGB (\cmark)    &                           & $0.0167\ (\pm\ 0.0464)$ & $0.1361\ (\pm\ 0.0100)$ & $0.0848\ (\pm\ 0.0445)$ & $\textbf{374.12}\ (\pm\ 6.17)$ \\
MW (months) (\cmark)  &                         & $0.0363\ (\pm\ 0.1880)$ & $0.5159\ (\pm\ 0.0908)$ & $0.3699\ (\pm\ 0.2116)$ & -- \\
MW (quarters) (\cmark) &                         & $0.1604\ (\pm\ 0.1612)$ & $0.1873\ (\pm\ 0.0270)$ & $0.1409\ (\pm\ 0.0641)$ & -- \\

\midrule
CL (\xmark)     & \multirow{6}{*}{Beta}    & $0.1405\ (\pm\ 0.0417)$ & $0.2182\ (\pm\ 0.0220)$ & $0.1426\ (\pm\ 0.0387)$ & -- \\
COX (\cmark)    &                           & $\textbf{0.0121}\ (\pm\ 0.0757)$ & $0.1625\ (\pm\ 0.0140)$ & $\textbf{0.1065}\ (\pm\ 0.0554)$ & $416.19\ (\pm\ 7.72)$ \\
NN (\cmark)     &                           & $0.0270\ (\pm\ 0.1088)$ & $0.1687\ (\pm\ 0.0189)$ & $0.1098\ (\pm\ 0.0556)$ & $416.75\ (\pm\ 8.60)$ \\
XGB (\cmark)    &                           & $0.0124\ (\pm\ 0.0629)$ & $\textbf{0.1612}\ (\pm\ 0.0122)$ & $0.1173\ (\pm\ 0.0552)$ & $\textbf{416.56}\ (\pm\ 7.48)$ \\
MW (months)   (\cmark)&                    & $0.0988\ (\pm\ 0.1047)$ & $0.5666\ (\pm\ 0.1191)$ & $0.3994\ (\pm\ 0.2462)$ & -- \\
MW (quarters) (\cmark)&                    & $0.2551\ (\pm\ 0.2565)$ & $0.2295\ (\pm\ 0.0505)$ & $0.1757\ (\pm\ 0.0838)$ & -- \\

\midrule
CL (\xmark)     & \multirow{6}{*}{Gamma}   & $0.2314\ (\pm\ 0.0399)$ & $0.2593\ (\pm\ 0.0339)$ & $0.2315\ (\pm\ 0.0399)$ & -- \\
COX (\xmark)    &                           & $\textbf{0.0465}\ (\pm\ 0.0485)$ & $\textbf{0.1507}\ (\pm\ 0.0147)$ & $\textbf{0.0912}\ (\pm\ 0.0467)$ & $408.00\ (\pm\ 6.06)$ \\
NN (\cmark)     &                           & $0.0478\ (\pm\ 0.0965)$ & $0.1655\ (\pm\ 0.0420)$ & $0.1093\ (\pm\ 0.0620)$ & $408.02\ (\pm\ 8.94)$ \\
XGB (\cmark)    &                           & $0.0799\ (\pm\ 0.0532)$ & $0.1536\ (\pm\ 0.0235)$ & $0.1049\ (\pm\ 0.0441)$ & $\textbf{407.01}\ (\pm\ 5.80)$ \\
MW (months) (\xmark)  &                     & $0.4092\ (\pm\ 0.1666)$ & $0.7368\ (\pm\ 0.1534)$ & $0.5876\ (\pm\ 0.2397)$ & -- \\
MW (quarters) (\xmark) &                    & $0.5067\ (\pm\ 0.2339)$ & $0.3601\ (\pm\ 0.0472)$ & $0.3203\ (\pm\ 0.0882)$ & -- \\

\midrule
CL (\xmark)     & \multirow{6}{*}{Delta}   & $0.1256\ (\pm\ 0.0355)$ & $0.2987\ (\pm\ 0.0213)$ & $0.1294\ (\pm\ 0.0324)$ & -- \\
COX (\cmark)    &                           & $0.0274\ (\pm\ 0.0433)$ & $0.2115\ (\pm\ 0.0235)$ & $\textbf{0.1064}\ (\pm\ 0.0683)$ & $394.21\ (\pm\ 6.77)$ \\
NN (\cmark)     &                           & $0.0683\ (\pm\ 0.0648)$ & $0.2143\ (\pm\ 0.0280)$ & $0.1231\ (\pm\ 0.0689)$ & $395.76\ (\pm\ 6.97)$ \\
XGB (\cmark)    &                           & $\textbf{0.0233}\ (\pm\ 0.0488)$ & $\textbf{0.1668}\ (\pm\ 0.0187)$ & $0.1219\ (\pm\ 0.0501)$ & $\textbf{376.80}\ (\pm\ 6.59)$ \\
MW (months)  (\xmark)  &                    & $0.1289\ (\pm\ 0.1179)$ & $0.7670\ (\pm\ 0.0867)$ & $0.5329\ (\pm\ 0.3327)$ & -- \\
MW (quarters)  (\xmark) &                   & $0.3177\ (\pm\ 0.1863)$ & $0.3766\ (\pm\ 0.0442)$ & $0.2839\ (\pm\ 0.1272)$ & -- \\

\midrule
CL (\xmark)     & \multirow{6}{*}{Epsilon} & $\textbf{0.0036}\ (\pm\ 0.0259)$ & $\textbf{0.1182}\ (\pm\ 0.0100)$ & $\textbf{0.0416}\ (\pm\ 0.0055)$ & -- \\
COX (\xmark)    &                           & $0.0078\ (\pm\ 0.0422)$ & $0.1329\ (\pm\ 0.0153)$ & $0.0817\ (\pm\ 0.0409)$ & $350.45\ (\pm\ 5.28)$ \\
NN (\xmark)     &                           & $0.0162\ (\pm\ 0.0510)$ & $0.1354\ (\pm\ 0.0160)$ & $0.0869\ (\pm\ 0.0412)$ & $351.19\ (\pm\ 5.25)$ \\
XGB (\xmark)    &                           & $0.0252\ (\pm\ 0.0903)$ & $0.1476\ (\pm\ 0.0563)$ & $0.0941\ (\pm\ 0.0557)$ & $\textbf{350.17}\ (\pm\ 5.23)$ \\
MW (months) (\cmark)  &                     & $0.0592\ (\pm\ 0.1018)$ & $0.4629\ (\pm\ 0.0734)$ & $0.3327\ (\pm\ 0.1797)$ & -- \\
MW (quarters)(\cmark) &                     & $0.1349\ (\pm\ 0.1255)$ & $0.1668\ (\pm\ 0.0252)$ & $0.1245\ (\pm\ 0.0542)$ & -- \\

\midrule
CL (\xmark)         & \multirow{6}{*}{Zeta} & $0.1415\ (\pm\ 0.0346)$ & $0.1968\ (\pm\ 0.0250)$ & $0.1416\ (\pm\ 0.0347)$ & -- \\
COX (\xmark)        &                       & $\textbf{0.0935}\ (\pm\ 0.0922)$ & $0.1738\ (\pm\ 0.0543)$ & $\textbf{0.0894}\ (\pm\ 0.0611)$ & $317.80\ (\pm\ 15.40)$ \\
NN (\xmark)         &                       & $0.1522\ (\pm\ 0.1147)$ & $0.2000\ (\pm\ 0.0612)$ & $0.1328\ (\pm\ 0.0757)$ & $318.18\ (\pm\ 32.91)$ \\
XGB (\xmark)        &                       & $0.1024\ (\pm\ 0.1375)$ & $\textbf{0.1847}\ (\pm\ 0.1206)$ & $0.0951\ (\pm\ 0.1278)$ & $\textbf{244.05}\ (\pm\ 8.22)$ \\
MW (months)(\cmark) &                       & $0.1601\ (\pm\ 0.0986)$ & $0.4735\ (\pm\ 0.0459)$ & $0.1182\ (\pm\ 0.0437)$ & -- \\
MW (quarters) (\cmark) &                    & $0.1483\ (\pm\ 0.0498)$ & $0.2118\ (\pm\ 0.0384)$ & $0.1364\ (\pm\ 0.0404)$ & -- \\
\bottomrule
\end{tabular}
\caption{\label{tab:resultsSIM} Results of our data application on the simulated data sets. For each model (column one) and each scenario (column two), we show the average performance metrics across the 20 simulations (columns three to four). We highlight in bold the best score for each metric in each scenario.}
\end{table}

The results of the data application on simulated data are summarized in \cref{tab:resultsSIM}. In column one, we list the models included in the comparison for each scenario (column two). Similarly to \cref{tab:effectsforscenarios}, we denote with a check mark or an x mark if the model assumptions are satisfied or not. For each scenario and each performance metrics, we higlighlight in bold the best performing model. Let us first consider the comparison with CL in terms of $R^{\text{TOT}}$ $R^{\text{cell-wise}}$ and $R^{\text{cal-wise}}$. 
Our data application shows that CL is outperforming the individual models in scenario Alpha and in scenario Epsilon according to the $R^{\text{TOT}}$, the $R^{\text{cell-wise}}$ and the $R^{\text{cal-wise}}$. As expected, in scenario Epsilon, breaking down the proportionality assumption, the CL is better than the other models. Interestingly, when we inspect the most complex scenarios (Beta, Gamma, and Delta) our approach (COX, NN, and XGB) provides better scores than the CL model. Notably, we observe that COX and XGB are consistently the best performing models. Notwithstanding the intensive hyper parameters tuning that we performed, NN seem to perform worse than COX in most scenarios. In scenario Delta, where we introduce a seasonality effect we find that according to all the proposed scores XGB is the best model in terms of $R^{\text{TOT}}$ and $R^{\text{cell-wise}}$.
The $R^{\text{TOT}}$, $R^{\text{cell-wise}}$ and $R^{\text{cal-wise}}$ are a good benchmark to compare our approach to the CL but they are not proper scoring rules. Conversely, we can take the CRPS as the main criterion to rank our models. In the scenarios we considered, the CRPS indicates that XGB is outperforming NN and COX. Interestingly, while in scenarios Alpha, Beta, Gamma, and Epsilon the XGB (average) CRPS is close to COX, in scenario Delta we observe a major drop in the CRPS going from COX to XGB. 

In scenario Zeta, where we have the continuous feature \texttt{property\_value}, we find that COX and XGB obtain similar scores and outperform the other models that we included in our study.
Let us now compare our models to the MW approach using the metrics discussed above. In our simulations, the MW approach yields results comparable to the CL model in terms of $R^{\text{TOT}}$ since the two models have similar assumptions. However, the high $R^{\text{cell-wise}}$ and $R^{\text{cal-wise}}$ scores indicate that the model fails to capture the correct dynamics of the reporting delays. The MW model is particularly outperformed by the other models in scenarios Gamma and Delta, where an accident date effect on the reporting delay is present and the model assumptions are broken. Also in scenario Zeta, where the model assumptions are satisfied, the model is outperformed by the CL and our models. In both scenario Beta and scenario Zeta, where the claims composition evolves over time, we observe that the assumptions underlying the MW model are satisfied. Unlike the classical CL method, which does not incorporate claim-level features and thus mixes different types of claims, the MW model utilizes feature-dependent development factors. This allows it to distinguish between claim types, explaining its superior performance in scenario Beta. However, in scenario Zeta, where a continuous covariate is introduced, the performance of MW seems to be worse than CL. This is likely due to the discretization of the continuous covariate in the MW framework.

\section{Data application on real data}
\label{sec:dataapplicationrd}

In this section we show a case study based on a real data set from a large Danish insurance company. We had at disposal the company complete reserving data from $2012$ to $2022$ for a short tailed personal line product. Two categorical features are associated to each (claim \texttt{claim\_type} and \texttt{coverage\_key}). The complete description of the data set is available in \cref{tab:codandata}.

\begin{table}[ht]
\centering
\begin{tabular}{l|l}
Features & Description\\ \hline
\texttt{Claim\_number} & Policy identifier.\\
 \texttt{claim\_type} $\in  \left\{1, \ldots, 20 \right\}$ & Type of claim. \\
 \texttt{coverage\_key} $\in  \left\{1, \ldots, 16 \right\}$& Type of coverage. \\
\texttt{AM} & Accident month.\\
\texttt{CM} & Calendar month of report.\\
\texttt{DM} & Development month.\\
\texttt{incPaid} & Incremental paid amount. We will not use this information in the manuscript. \\
 \hline
  \end{tabular}
  \caption{\label{tab:codandata} Description of the real data set, available from a large Danish non-life insurer}. 
\end{table}

We decided to split the data into $6$ chunks of $5$ consecutive years each. The splits are reported in \cref{tab:splitsdata}, together with the corresponding data size. 
For each of these splits, we fit our models on the first $4$ years (train data) and score our models on the fifth year (test data).

\begin{table}[ht]
\centering
\small
\begin{tabular}{|c|c|c|c|c|c|}
 & Training & Scoring & Train data size (observations number)& Test data size (observations number) & Mean Reporting Delay\\ 
  \hline
Split 1 & 2012-2016 & 2017&129381 & 1474& 1.62 \\ 
Split 2 & 2013-2017 & 2018&133367 & 1553& 1.62 \\
Split 3 & 2014-2018 & 2019&130405 & 1566& 1.63 \\
Split 4 & 2015-2019 & 2020&122842 & 1394& 1.63 \\
Split 5 & 2016-2020 & 2021&110884 & 1225& 1.61 \\
Split 6 & 2017-2021 & 2022&105618 & 1069 & 1.55 \\
\hline
\end{tabular}
\caption{\label{tab:splitsdata} For each split (column one) we reported the accident periods we used to train the models (column two), and the year we use for scoring (column three). In columns four and five we show the train and test data size. In column six we provide the mean reporting delay from accident.}
\end{table}

Similarly to the case study on the simulated data, we compare our results with the chain ladder model (CL) fitted on a quarterly grid and the MW model fitted on a monthly and yearly grid. We fitted our models on a monthly grid, according to the procedure we described in the manuscript. In \cref{tab:splitsdata}, we show the mean reporting delay in each data split. Most of the notifications occurred in the early development periods and the mean reporting delay is between one and two months. 

\begin{figure}[ht] 
    \centering
    \includegraphics[width=0.9\linewidth]{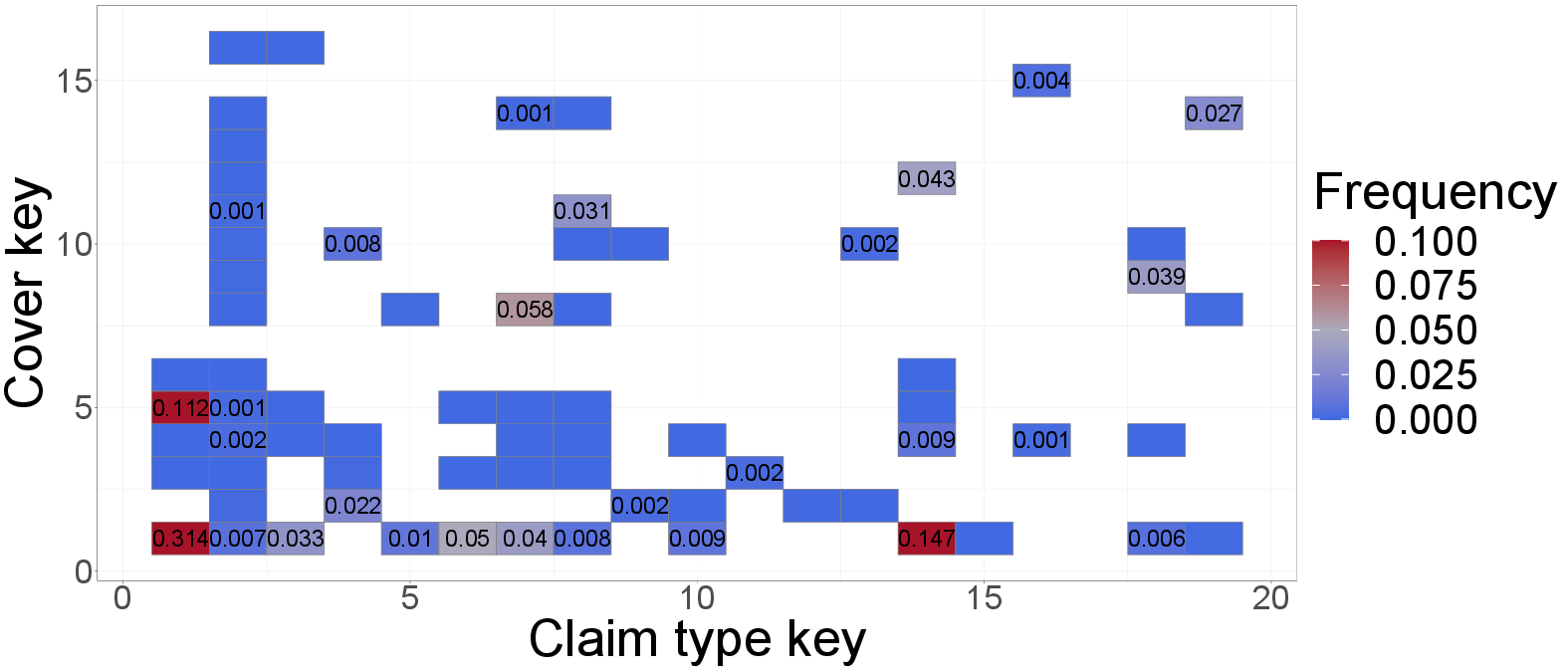}    
  \caption{\label{fig:tileplot} Relative frequency on the complete data ($2012-2021$). We do not observe all the combinations of features and most of the data show \texttt{claim\_type\_key} 1, and \texttt{cover\_key} 1 or \texttt{cover\_key} 5.}

\end{figure}

From \cref{fig:tileplot} we can see that not all the combinations of the two categorical features (\texttt{cover\_key} and \texttt{claim\_type\_key}) were observed (see \autoref{fig:tileplot}). Most of the data belong to the combinations \texttt{claim\_type\_key 1} with \texttt{cover\_key 1} and  \texttt{cover\_key 5} and \texttt{claim\_type\_key 14} with \texttt{cover\_key 1}.

\subsection{Results}

In \cref{tab:resultsRD} we show the average performance of our models  on the real data splits described in \cref{tab:splitsdata}.  

Interestingly, from the results of the $R^{\text{TOT}}$, $R^{\text{cell-wise}}$ and $R^{\text{cal-wise}}$ (columns three to five) we see that in general we have an improvement in the models performance using COX, NN, or XGB compared to the CL, especially in terms of $R^{\text{TOT}}$. Among our models, COX obtained the minimum CRPS score. In our application, the NN model obtained the best score in terms of $R^{\text{TOT}}$. Conversely, the CL obtained the lowest  $R^{\text{cell-wise}}$ and $R^{\text{cal-wise}}$. The MW models are outperformed from our models and the CL in all the inspected metrics. In this setting, the chain ladder is expect to perform very well as the data are reported in the early development months and, from the authors experience on this data, there are few reports in the late development periods. Notably, in this setting of quick settlement, our model exhibits performance comparable to that of the CL approach. 

\begin{table}[H]
\centering
\begin{tabular}{l c c c c}
\toprule
Model & $\texttt{EI}^{\texttt{R}}$ & $R^{\text{cell-wise}}$ & $R^{\text{cal-wise}}$ & CRPS (average) \\
\midrule
CL             & $0.0095\ (\pm\ 0.1353)$ & $\textbf{0.2174}\ (\pm\ 0.0248)$ & $\textbf{0.1345}\ (\pm\ 0.0420)$ & -- \\
COX            & $0.0099\ (\pm\ 0.1255)$ & $0.2403\ (\pm\ 0.0564)$ & $0.1896\ (\pm\ 0.0840)$ & $\textbf{5.92}\ (\pm\ 1.31)$ \\
NN             & $\textbf{0.0049}\ (\pm\ 0.1313)$ & $0.2449\ (\pm\ 0.0527)$ & $0.1930\ (\pm\ 0.0827)$ & $5.96\ (\pm\ 1.29)$ \\
XGB            & $0.0254\ (\pm\ 0.1271)$ & $0.2424\ (\pm\ 0.0556)$ & $0.2003\ (\pm\ 0.0878)$ & $6.01\ (\pm\ 1.28)$ \\
MW (monthly)   & $0.0291\ (\pm\ 0.7255)$ & $0.5031\ (\pm\ 0.2666)$ & $0.2591\ (\pm\ 0.1713)$ & -- \\
MW (quarterly) & $0.2629\ (\pm\ 0.8674)$ & $0.3694\ (\pm\ 0.1655)$ & $0.2194\ (\pm\ 0.1409)$ & -- \\
\bottomrule
\end{tabular}
\caption{\label{tab:resultsRD}  Results on the case study on real data. For each model (column one) we show the average results across the different data splits of $R^{\text{TOT}}$, $R^{\text{cell-wise}}$, $R^{\text{cal-wise}}$, and CRPS (columns two to four). The $R^{\text{cal-wise}}$ is presented quarterly. For each performance metric and each split, the best average score is highlighted in boldface.}
\label{tab:results_case_study}
\end{table}

\section{Concluding remarks}

Based on the work of \cite{miranda2013continuous} and \cite{hiabu17},
we introduced a survival analysis framework to use machine learning techniques to estimate development factors that can depend on accident date and other features. The approach presented in this paper has been developed with the  aim to give higher accuracy in cases where more information in form of individual claims data is available while at the same time conserving the structure reserving actuaries are used to.
Our extensive simulation study suggests that our methodology does indeed seem to work well.
In this paper, we have only considered the prediction of IBNR counts and an obvious next step is to integrate our methodology into a wider framework to estimate the outstanding claims amount.
It could for example be interesting to merge our approach with recent RBNS prediction methods like \cite{crevecoeur2022hierarchical} or \cite{lopez21}. Lastly, it could be interesting to extend our work beyond point predictions towards quantifying uncertainty.

\section{Supplementary material}

The code at \href{https://github.com/gpitt71/resurv-replication-code.git}{gpitt71/resurv-replication-code} complements the results of this manuscript and can be used to replicate the case study of \cref{sec:dataapplicationsim}. The GitHub folder \texttt{resurv-replication-code}, was registered with a unique \href{https://zenodo.org/records/10419197}{ Zenodo DOI}. The code to obtain the plots that we included in the manuscript can be found in the package vignette \texttt{Manuscript replication material} of \href{https://github.com/edhofman/ReSurv}{edhofman/ReSurv}. Also the GitHub folder \texttt{ReSurv}, was registered with a unique \href{https://zenodo.org/records/10418822}{Zenodo DOI}. 

\section*{Acknowledgements}

Gabriele Pittarello is funded by the Novo Nordisk Foundation grant NNF23OC0084961.

\section*{Disclosure Statement}

No potential conflict of interest was reported by the authors. ChatGPT 4.0 and later versions were used for interactive online search with the LLM-enhanced search engines.

\bibliography{manuscript}  

\appendix

\section{Data pre-processing}
\label{appendix:datapp}

In this section we provide some further details on the data-preprocessing of categorical and continuous covariates that we implemented in our paper.

Let us consider some categorical covariate $X^\nu \in \mathbb{X}$  with $ \mathbb{X}=\left\{ X^{\nu,1}, \ldots, X^{\nu,L} \right\}$. We denote the dummy encoding function as $D:\mathbb{X} \rightarrow \left\{0,1\right\}^L$. For some individual $i$ in our data set, we will observe some record of $X^\nu$, $x_i^\nu \in \mathbb{X}$ that can be represented as 

$$D(x_i^\nu)= \left[I(x=X^{\nu,1}), \ldots,  I(x=X^{\nu,L})\right].$$ 

Let us now consider some continuous covariate $X^c \in \mathbb{R}$. In our data set, we will have the observations $x^c_1, \ldots, x^c_n$ of $X^c$. Let us define $x_{\min} = \min_{1 \leq i \leq n} x_i, \quad x_{\max} = \max_{1 \leq i \leq n} x_i$. We use the minimum-maximum function $MinMax: \mathbb{R} \rightarrow \left[0,1\right]$. Each of the observed records $x^c_1$ of the continuous features is transformed as 

$$MinMax(x^c_i)=2\frac{x^c_i-x_{\min}}{x_{\max}-x_{\min}}.$$

\section{Bayesian optimization of machine learning algorithms}
\label{appendix:bayescv}

In this paper we showed an extended analysis of $5$ simulated scenarios and a case study where we use machine learning to process large data sets and catch complex interactions in the data. Machine learning algorithms are very sensitive to the parameters and hyper parameters choice (\citeNP[p.~219]{hastie09}, \citeNP[]{snoek12}). 
In this section, we provide general details about the strategy that we used for the hyper parameters selection of NN and XGB. Optimizing the algorithms over many data sets might lead to protracted computational times, \cite{jones01}. As a solution, we use the Bayesian optimization procedure described in \citeA{snoek12}. While grid searches can be cumbersome for big searches, using the approach in \cite{snoek12} we use the information from prior model evaluations to guide the optimal parameters search. Notably, Bayesian optimization methods have shown to be well-performing in challenging optimization problems \cite{jones01}. 

Let us consider the data  $\left\{(x_i, y_i)\right\}_{i=1, \ldots, n}$. In this setting, the functional relationship between input and output $h: X \rightarrow \mathbb{R}$ is modelled as: 

\begin{align*}
    y_n \sim \mathcal{N}\left(h(x_n), \nu\right),
\end{align*}

with $\nu$ being the variance of noise introduced into the function observations. Furthermore, let us assume that the observations $h\left(x\right)$ are drawn from a Gaussian prior.
\newline

Under the Gaussian Process prior, we get a posterior function (i.e.,  $a$ acquisition function) $a: X \rightarrow \mathbb{R}^+$ that depends on the model solely through its predictive mean function $\mu(x; \left\{(x_i, y_i)\right\}_{i}, \theta )$ and predictive variance function $\sigma^2(x; \left\{(x_i, y_i)\right\}_{i}, \theta )$. Conversely, the acquisition function depends on the previous observations, as well as the Gaussian Process hyperparameters $a(x; \left\{(x_i, y_i)\right\}_{i}, \theta )$. In order to avoid an involved notation we denote

\begin{itemize}
    \item $a(x; \left\{(x_i, y_i)\right\}_{i}, \theta )$ as $a\left(x\right)$
    \item $\mu(x; \left\{(x_i, y_i)\right\}_{i}, \theta )$ as $\mu\left(x\right)$
    \item $\sigma^2(x; \left\{(x_i, y_i)\right\}_{i}, \theta )$ as $\sigma^2\left(x\right)$
\end{itemize}

There are different definitions for the acquisition function \cite{snoek12}, we choose

\begin{align}
\label{eq:acquifun}
    a\left(x\right)=\left(\mu(x)-h_{\max }\right) \Phi\left(\frac{\mu(x)-h_{\max }-\xi}{\sigma(x)}\right)+\sigma(x) \phi\left(\frac{\mu(x)-h_{\max }-\xi}{\sigma(x)}\right),
\end{align}

where

\begin{itemize}
    \item  $h_{\max }$ is the current maximum value obtained from sampling.
    \item $\Phi$ is the standard normal cumulative density function.
    \item $\phi$ is the standard normal probability density function.
    \item $\xi$ is an exploration parameter \cite{ParBayesianOptimization}.
\end{itemize}

This approach for Bayesian optimization is described with the following steps:

\begin{enumerate}
    \item Set the parameters to an initial value.
    \item Fit the Gaussian process.
    \item Find the parameters that maximize the acquisition function.
    \item Score the parameter.
    \item Repeat steps 2-4 until some stopping criteria is met \cite{snoek12}.
\end{enumerate}

A thorough description can be found in \cite{snoek12}. We show the hyperparameters we inspected for each model in \autoref{tab:hpselection}. 

\begin{table}[ht]
\centering
\begin{tabular}{l|l|l|p{7cm}}
Model & Hyperparameter & Range & Description \\ \hline
\multirow{7}{*}{NN} 
& \texttt{num\_layers} & $[2,10]$ & Defines the depth of the network. \\
& \texttt{num\_nodes} & $[2,10]$ & Determines the width of each layer. \\
& \texttt{optim} & $[1,2]$ & Algorithm to update model weights based on the loss gradient. \\
& \texttt{activation} & $[1,2]$ & Introduces non-linearity in the network. \\
& \texttt{lr} & $[.005,0.5]$ & Controls step size during objective optimization. \\
& \texttt{xi} & $[0,0.5]$ & Partial likelihood elastic regularization. \\
& \texttt{eps} & $[0,0.5]$ & Partial likelihood elastic regularization. \\\hline

\multirow{6}{*}{XGB} 
& \texttt{eta} & $[0,1]$ & Controls step size during objective optimization. \\
& \texttt{max\_depth} & $[0,25]$ & Maximum depth of trees. Controls model complexity. \\
& \texttt{min\_child\_weight} & $[0,50]$ & Assures no final node is too small. \\
& \texttt{subsample} & $[0.1,1]$ & Fraction of samples used for training each tree. \\
& \texttt{lambda} & $[0,50]$ & L2 regularization term on weights. \\
& \texttt{alpha} & $[0,50]$ & L1 regularization term on weights. \\\hline

\multirow{2}{*}{MW} 
& Number of hidden nodes & $[1,10]$ & Number of hidden nodes in the hidden layers. \\
& Number of hidden layers & $[1,2]$ & Number of hidden layers. \\\hline
\end{tabular}
\caption{\label{tab:hpselection} The range of hyperparameters we inspected. We set the same ranges for the $5$ simulations and the case study on the real data. } 
\end{table}

Below we disclose the computational times we required for fitting the parameters combinations on the five simulated scenarios (\autoref{tab:comptsim}) and the real data (\autoref{tab:comptrd}).

\begin{table}[ht]
\centering
\begin{tabular}{|l|l|c|c|}
 Model & Scenario & Hyperparameter selection & Model fit \\ 
  \hline
COX & \multirow{3}{1cm}{Alpha} &  & 3.20 \\ 
NN &  & 52.57 & 149.31 \\ 
XGB &  & 3.19 & 16.77 \\ 
   \hline
COX & \multirow{3}{1cm}{Beta} &  & 2.49 \\ 
NN &  & 44.00 & 102.13 \\ 
XGB &  & 2.83 & 15.66 \\ 
   \hline
COX & \multirow{3}{1cm}{Gamma} &  & 2.25 \\ 
NN &  & 68.72 & 204.58 \\ 
XGB &  & 5.10 & 21.91 \\ 
   \hline
COX & \multirow{3}{1cm}{Delta} &  & 2.23 \\ 
NN &  & 75.05 & 132.79 \\ 
XGB &  & 8.72 & 19.61 \\ 
   \hline
COX & \multirow{3}{1cm}{Epsilon} &  & 1.94 \\ 
NN &  & 66.33 & 109.64 \\ 
XGB &  & 3.51 & 11.43 \\ 
   \hline
\end{tabular}
\caption{\label{tab:comptsim} Average computational times in minutes, simulated scenarios. Hyperparameters selection is on 3-folded cross validation. Model fit includes development factor fitting.} 
\end{table}

\begin{table}[ht]
\centering
\begin{tabular}{|l|l|c|c|}
 Model & Split & Hyperparameter selection & Model fit \\ 
  \hline
COX & \multirow{3}{1cm}{Split 1} &  & 1.53 \\ 
  NN & & 88.28 & 27.18 \\ 
  XGB &  & 10.57 & 21.56 \\ 
   \hline
COX & \multirow{3}{1cm}{Split 2} &  & 1.37 \\ 
  NN &  & 93.30 & 32.41 \\ 
  XGB &  & 31.25 & 24.06 \\ 
   \hline
COX & \multirow{3}{1cm}{Split 3} &  & 0.94 \\ 
  NN & & 161.71 & 40.91 \\ 
  XGB & & 19.01 & 25.86 \\ 
   \hline
COX & \multirow{3}{1cm}{Split 4} &  & 1.80 \\ 
  NN &  & 107.15 & 38.28 \\ 
  XGB &  & 65.10 & 28.01 \\ 
   \hline
COX & \multirow{3}{1cm}{Split 5} &  & 1.74 \\ 
  NN &  & 115.49 & 20.78 \\ 
  XGB &  & 38.12 & 26.80 \\ 
   \hline
COX & \multirow{3}{1cm}{Split 6} &  & 1.24 \\ 
  NN &  & 87.51 & 19.99 \\ 
  XGB & & 35.45 & 27.38 \\ 
   \hline
\end{tabular}
\caption{\label{tab:comptrd} Computational times in minutes, real data. Hyperparameters selection is on 3-folded cross validation. Model fit includes development factor fitting.} 
\end{table}

\section{Details on Wüthrich (2018)}
\label{appendix:mw}

Under the regularity conditions in \citeA{wuthrich18}, for $j=0,\ldots, J-1$ the MW approach proposes a model for $f^*_{j}(x)$ feature-dependent development factors of development triangles with non-zero entries

$$
\mathbb{E}\left[C_{k, j}(x) \mid \mathcal{F}_{k+j-1}\right]=f^*_{j-1}(x) C_{k, j-1}(x),
$$

where $\mathcal{F}_{t}$ represents the filtration at time $t$. An estimator $\hat f^*_{j-1}(x)$ of the $f^*_{j-1}(x)$ development factors is obtained by minimizing the weighted squared loss function

\begin{align}
\label{eq:mw_lkh}
\mathcal{L}_j^{MW} \propto \sum_{k=1}^{J-j} \sum_{x: C_{k, j-1}(x)>0} C_{k, j-1}(x)\left(\frac{C_{k, j}(x)}{C_{k, j-1}(x)} - f^*_{j-1}(x)\right)^2,
\end{align}

where $f^*_{j-1}(x): x \rightarrow \mathbb{R}^+$ is modeled as the prediction target of a neural network. In \citeA{wuthrich18}, the authors use one hidden layer with the following architecture 

$$f^*_{j-1}(x)= \text{exp}\left( \text{tanh}\left(x^T w^{(i)} \right) w^{(h)} \right),$$

where $w^{(i)} \in \mathbb{R}^p$ and $w^{(h)} \in \mathbb{R}^{p b}$ are the neural network parameters obtained minimising \cref{eq:mw_lkh}. In our notation, $b$ is the number of hidden nodes. 
\newline

For selecting the depth of the neural network, the activation function at the hidden layer, and the activation function at the output layer of the neural network used in the MW model, we adopted the Bayesian model for hyperparameters selection that we also used for our individual models from \citeA{snoek12} described in \cref{appendix:bayescv}. 
In the MW approach, the choice of a strictly positive activation function is necessary to guarantee the development factors to be positive, while they could yet be estimated lower than one implying negative payments. 

For $j+k>K-1$ the estimated $\hat f^*_{j-1}(x)$, with $j=0,\ldots, J-1$  development factors are then used for prediction of the lower triangle of non-zero claims in a similar fashion to our proposal in \cref{ss:predictions}.  In contrast to our proposal, the MW approach also requires a separate model to handle cases where $C_{k,K-k-1}(x) = 0$, as described in Section 4 of \citeA{wuthrich18}. Notably, it is in not clear how one can describe a reasonable data generating mechanism that fits all model assumptions.

By comparing \cref{eq:mw_lkh} to the partial log-likelihood that we minimize, one can identify a key distinction between our approach and the MW approach. Specifically, while our method estimates the development factors $f_{k,j}(x)$ using individual observations, the MW approach relies on aggregating the data into feature-dependent triangles for the estimation of the development factors $f^*_{j-1}(x)$. The feature-dependent development triangles can have different granularities and in our paper, we will inspect the results for the MW model both on a monthly and a quarterly grid.

Using the survival analysis approach that we propose, it is also possible to take into account the accident period as an effect on the hazard for additional flexibility compared to the standard chain-ladder framework. Using the model in \cref{eq:mw_lkh}, this would only be feasible for in-sample estimation but would require extrapolating an accident period effect for out-of-sample predictions.

\section{Scenarios simulation}
\label{appendix:simulation}

In \autoref{sec:dataapplicationsim} we illustrated the steps that we followed to generate the five simulated scenarios. We also mentioned that the parametrization of $\phi(x,u;\theta)$ changes for the different scenarios. In this section we want to provide extra details on the parameters that we used in the simulation phase. For the data generation we will only need two modules from the \texttt{SynthETIC} package, i.e. the number of claims occurring every accident date and the reporting delay. In every scenario for both claim types, the rate of claims occurrence is $.2$. In scenarios Alpha, Gamma, Delta, and Epsilon the individuals at risk in the portfolio are $200$ in each accident date. In scenario Beta the individuals for \texttt{claim\_type 1} are decreasing. 

In \autoref{tab:rtfwd1} we report the parameters that we used to simulate from the RTFWD distribution in \cite{teamah19}. We recall that the RTFWD has a four parameter structure $\nu, \pi, \xi, k$, and is defined with $0<t \leq b$ with cumulative distribution function

$$
F(t) = \exp(-\pi^{\nu}\xi^{\nu k}(t^{-\nu k}-b^{-\nu k})).
$$

\begin{table}[ht]
\centering
\begin{tabular}{l|c|c|c|c}
 Scenario & $\nu$ & $\pi$ & $k$ & $\xi_0$ \\ \hline
 Alfa, Beta, Gamma, Delta &  0.5 & 60 & 1 & 0.1  \\
 Epsilon & 0.5 & $60+(34.5387 \; I(\texttt{claim\_type }=0)  + 58.6803 \;I(\texttt{claim\_type } =0))$ & 1 & 0.1\\
 \hline
  \end{tabular}
\caption{\label{tab:rtfwd1} The RTFWD distribution parameters (columns two to four) for the different scenarios (column one).} 
\end{table}

In \autoref{tab:abgdepars} we reported the parameters that we used for the simulation of the hazard. In the scenarios Alpha, Beta, Gamma, and Delta the reverse time hazard has the form

$$
\alpha(t|\text{\texttt{AD}},\text{\texttt{claim\_type}})= \alpha_0(t)e^{\phi \left(\text{\texttt{AD}},\text{\texttt{claim\_type}}; \theta \right)}.
$$

In column two of Panel A of \autoref{tab:abgdepars} we show the baseline $\alpha_0(t)$. We highlight that $\alpha_0(t)$ is the same for the five scenarios. In column three of panel A we show the different effects of the features on the proportional risk $e^{\phi \left(\text{\texttt{AD}},\text{\texttt{claim\_type}}; \theta \right)}$ for the different scenarios (column 1).

In scenario Epsilon the hazard is not generated from a proportional model and it has the form:

$$
\alpha(t|\text{\texttt{AD}},\text{\texttt{claim\_type}})= \alpha_0(t|\text{\texttt{claim\_type}})(\exp(\phi \left(\text{\texttt{AD}},\text{\texttt{claim\_type}}; \theta \right))+f(\texttt{claim\_type})).
$$

Details on the simulation for this scenario are provided in Panel B of \autoref{tab:abgdepars}. In Panel C we show the coefficients values that we chose to simulate the proportional risk for the different scenarios (columns one to seven). 

\begin{table}[ht]
\centering
\caption{\label{tab:abgdepars} The parameters that we used for the reverse hazard simulation in the different scenarios. Baseline (column two) and effects on the proportional risk (column three) for the five scenarios (column one).  Let us define $I(\text{\texttt{AD}})=\text{\texttt{AD}}-30\% (\text{\texttt{AD}}-1)/30 \rfloor$} 

\begin{tabular}{l|c|c}
 Scenario &  $\alpha_0(t)$ & $\phi \left(\text{\texttt{AD}},\text{\texttt{claim\_type}}; \theta \right)$  \\ \hline
 Alpha & \multirow{4}{2cm}{$0.5\sqrt{0.2 t^{-1}} $} &$ \beta_0 I(\texttt{claim\_type } =0) + \beta_1 I(\texttt{claim\_type } =1)$ \\ \cline{1-1} \cline{3-3} 
Beta & &$  \beta_0 \;I(\texttt{claim\_type } =0) + \beta_1 \;I(\texttt{claim\_type } =1)$\\ \cline{1-1} \cline{3-3}
Gamma & &  \makecell[l]{$\beta_0\; \;I(\texttt{claim\_type } =0) +\beta_1\; \;I(\texttt{claim\_type } =1)  + \beta_2\; \;I(\texttt{claim\_type } =1)\sqrt{\text{\texttt{AD}}}$ }  \\ \cline{1-1} \cline{3-3}
Delta & & \makecell[l]{$\beta_0\,I(\texttt{claim\_type} = 0) + \beta_1\,I(\texttt{claim\_type} = 1) + \beta_3\,I(\texttt{AD} \in \{2, 3, 4\}) $\\$+ \beta_4\,I(\texttt{AD} \in \{5, 6, 7\}) + \beta_5\,I(\texttt{AD} \in \{8, 9, 10\}) + \beta_6\,I(\texttt{AD} \in \{11, 0, 1\})$}\\ \hline
\end{tabular}
\end{table}

\begin{table}[ht]
\centering
\begin{tabular}{l|c|c|c}
 Scenario & $f(\text{\texttt{claim\_type}})$&  $\alpha_0(t|\text{\texttt{claim\_type}})$ & $\phi \left(\text{\texttt{AD}},\text{\texttt{claim\_type}}; \theta \right)$\\ \hline
Epsilon &$0.5 \phi \left(\text{\texttt{AD}},\text{\texttt{claim\_type}}; \theta \right)$& $0.5\sqrt{0.1(2+f(x))) t^{-1}} $&$ 
\beta_0\;I(\texttt{claim\_type } =0) +\beta_1\;I(\texttt{claim\_type } =1)$\\
 \hline
\end{tabular}
\caption{Panel B: Baseline (column three), effects on $\phi \left(\text{\texttt{AD}},\text{\texttt{claim\_type}}; \theta \right)$ (column four) and $f(\text{\texttt{claim\_type}})$ for scenario Epsilon (column one).}
\end{table}

\begin{table}[H]
\centering
\begin{tabular}{c|c|c|c|c|c|c}
 $\beta_0$&  $\beta_1$ & $\beta_2$ &$\beta_3$ & $\beta_4$ & $\beta_5$ & $\beta_6$ \\ \hline
$1.1512$& $1.95601$ &$-0.021206$&$-0.3$&$0.4$&$-0.7$&$0.1$\\
 \hline
  \end{tabular}
\caption{Panel C: Features effects, $\beta_0, \ldots \beta_6 \in \mathbb{R}$}
\end{table}

In scenario \textbf{Zeta}, the delay from accident to report $t \in \mathbb{R}^+$ follows a Transformed Gamma distribution with parameters $s_1,s_2,o >0$ and density

\begin{equation}
f(t|s_1,s_2,o)=\frac{s_2 (t/o)^{s_1s_2} e^{-(t/o)^{s_2}}}{t \Gamma(s_1)}.
\end{equation}

The Transformed Gamma distribution parameters in this scenario are

     \begin{align*}
     s_1 &= 1 + \frac{1}{\texttt{property\_value}}. \\
     s_2 &= 1 - \frac{1 + \mathbb{I}\{\texttt{business\_use} = \text{Y}\}}{10}. \\
     o   &= 0.2.
 \end{align*}

Lastly, \texttt{property\_value} was generated from a lognormal distribution with logarithm mean $3.034513$ and logarithmic standard deviation $0.4087569$. As mentioned in the main text, to obtain the parameters of the distribution generating \texttt{property\_value}, we calibrated the lognormal distribution on the  house prices of the Boston Housing Prices data set as available from the \texttt{R} package \texttt{A3}. For obtaining the the \texttt{age} covariate, we sample from uniform with support between 50 and 55.

\section{More on our model output}
\subsection{Minimizing the log-likelihood}
\label{appendix:llminimization}

In this section we will show the hazard models average in-sample negative partial log-likelihood, i.e. the loss function we minimize during the models training computed on the data that we used to fit the models. In order to ease our notation, we will indicate the average negative log-likelihood in \autoref{eq:loglkh} as $l$.
The model with the minimum in-sample average negative partial log-likelihood is the model that fits best the training data. 
We train the COX model using all the individual data from calendar periods $\tau=1,\ldots, \mathcal{T}$. In the training phase of XGB and NN, the input data from calendar periods $\tau=1,\ldots, \mathcal{T}$ are further split into a main part for training and a smaller part for validation. The split is random and we use $80 \%$ of the data for training (the splitting percentage is selected with a rule of thumb). We will then report for XGB and NN the out-of-sample average negative log-likelihood measured on the remaining $20 \%$ of the input data. Comparing the in-sample likelihood and the out-of-sample likelihood will tell us whether a model is overfitting the data. Indeed, we expect that a) the models can be ordered in the same way based on their descending score in-sample and out-of-sample and b) the magnitude of the scores in-sample and out-of-sample is comparable.
In \cref{tab:lkhsim} we show the (average) negative log-likelihood averaged (over the $20$ simulations), for each model in each scenario. The results show that XGB and NN seem to best fit the in-sample data compared to COX. Furthermore, XGB is consistently providing a lower likelihood compared to the NN. A similar behavior is reflected in the out-of-sample data.

\begin{table}[ht]
\centering
\begin{tabular}{|l|l|c|c|}
 Model & Scenario & $l(\theta)$ (in-sample)  & $l(\theta)$ (out-of-sample) \\
  \hline
COX & \multirow{3}{1.1cm}{Alpha} & 9.238 & - \\ 
NN &  & 8.648 & 7.263\\ 
XGB &  & 8.634 & 7.249\\ 
\hline
COX & \multirow{3}{1.1cm}{Beta}  & 9.112 & - \\ 
NN &  & 8.512 & 7.127 \\ 
XGB &  & 8.502 & 7.118\\ 
\hline
COX & \multirow{3}{1.1cm}{Gamma}  & 9.062 & - \\ 
NN &  & 8.760 & 7.374 \\ 
XGB &  & 8.755 & 7.372\\
\hline
COX & \multirow{3}{1cm}{Delta}  & 9.199 & - \\ 
NN &  & 8.686 & 7.303 \\ 
XGB &  & 8.590 & 7.215\\
\hline
COX & \multirow{3}{1.1cm}{Epsilon}  & 9.121 & -\\ 
NN &  & 8.563 & 7.180\\ 
XGB &  & 8.556 & 7.175\\ 
   \hline
\end{tabular}
\caption{\label{tab:lkhsim} In each scenario, the average log-likelihood $l(\theta)$ is computed in each simulation on each simulated data set. For machine learning models each data set is split in training (in-sample, $80 \%$ of the data) and validation (out-of-sample, $20 \%$ of the data). The splitting percentage is determined as a rule of thumb. Here, we provide for each scenario the results of the average likelihood $l(\theta)$ (in-sample and out-of-sample) over the $20$ simulations. } 
\end{table}

A similar table can be provided for the real world data application \cref{tab:resultsRD}. 

\subsection{Modeling the survival function}

In the previous sections, we observed that our predictions rely on a discrete time framework meanwhile we defined the hazards in continuous time. In the simulated scenarios we have, for $t \geq 0 $, a closed form for the features dependent survival function ${S}(t|{X},U)$. We can compare the survival function with the survival function we estimated with our models, i.e. $\hat{S}(t|{X},U)$ in \cref{eq:estimatedsf}. Visually inspecting the survival function and comparing it in different scenarios for different models can help in understanding the models fit across different scenarios.  

\begin{figure}
\centering
 \includegraphics[width=\linewidth]{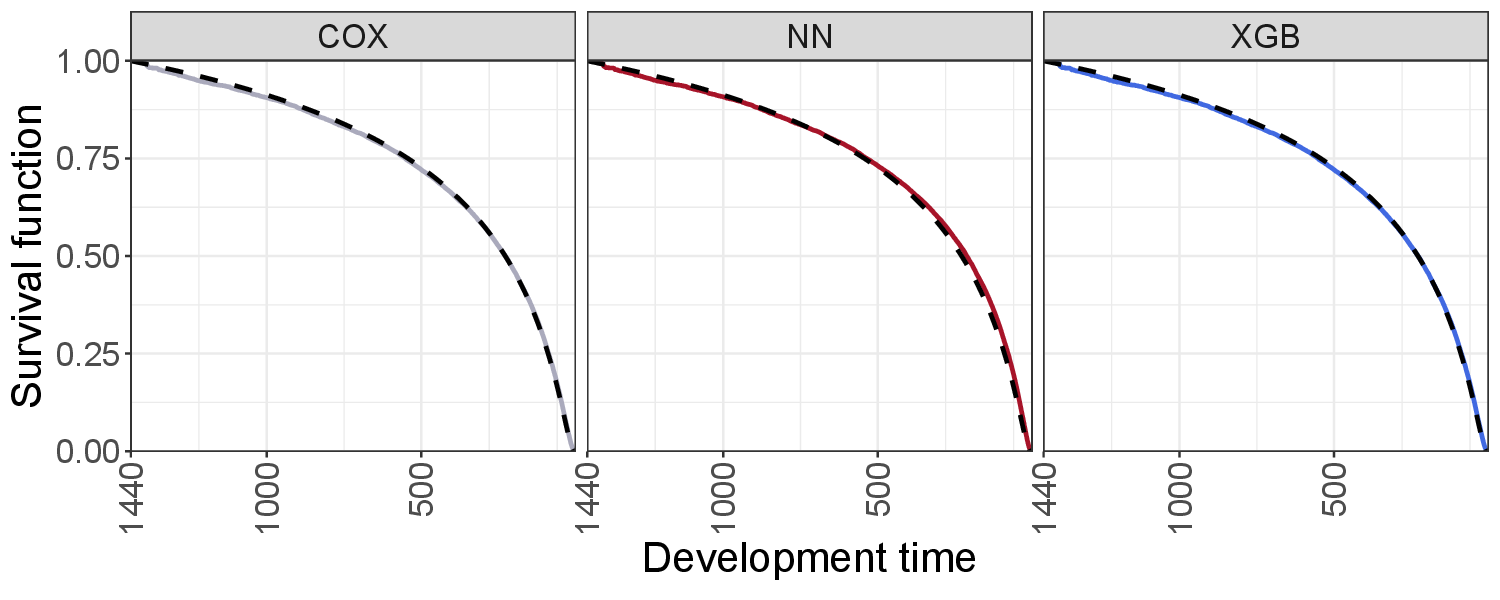}
 \caption{\label{fig:sfalpha} Scenario Alpha, \texttt{claim\_type 1} and \texttt{AD 13}. The true survival function (black dotted line) is compared to the fitted survival function with COX, NN, and XGB (left to right).}
 \end{figure}

In \cref{fig:sfalpha} we show the true survival function compared to the fitted survival function in scenario Alpha. It can be noticed that all the three models seem to behave consistently with respect to the true curve. Interestingly, even in the simplest modeling scenario, XGB and COX seem to be better than NN in catching the behavior of the survival function on the right tail. 

\begin{figure}
\centering
 \includegraphics[width=\linewidth]{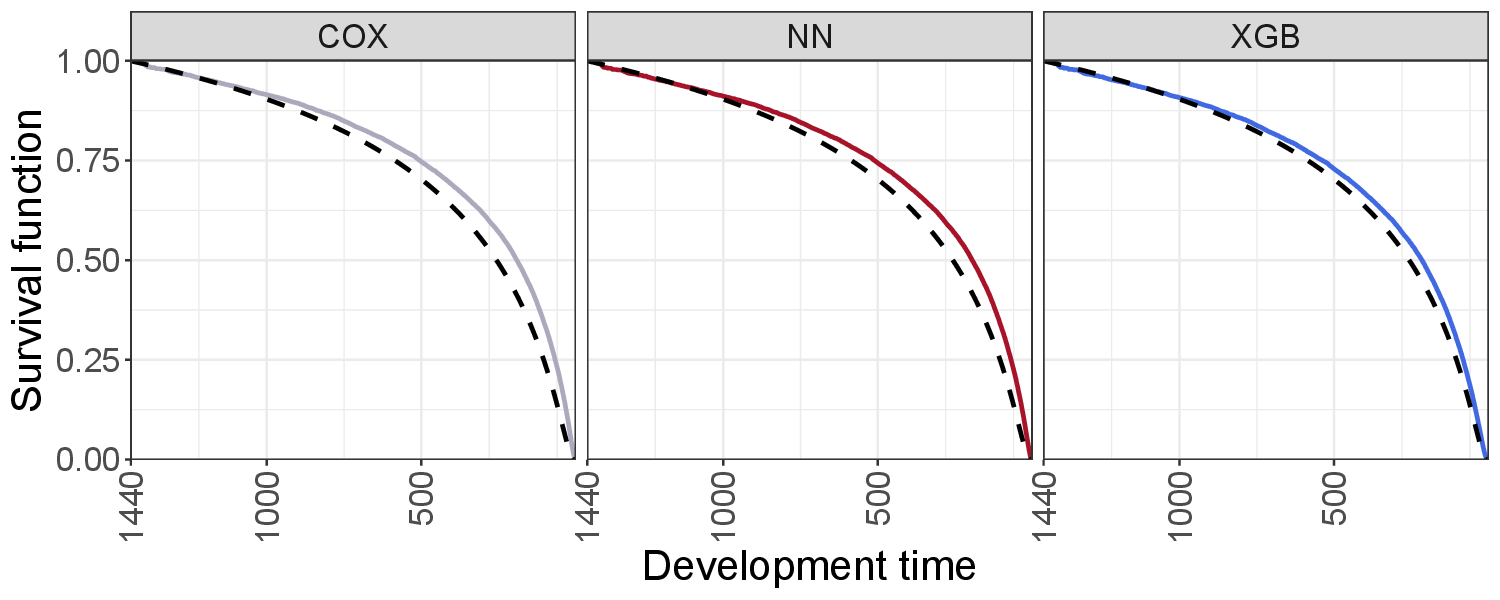}
 \caption{\label{fig:sfdeltaap691} Scenario Delta, \texttt{claim\_type 1} and \texttt{AD 691}. The true survival function (black dotted line) is compared to the fitted survival function with COX, NN, and XGB (left to right).}
 \end{figure}

The same plot can be shown for other scenarios. An interesting case is scenario Delta, where we introduced a seasonality effect dependent on the accident period, see again \cref{tab:effectsforscenarios}. In this section we first show the feature combination \texttt{claim\_type 1} and \texttt{AD 691} where the three models seem to behave similarly with XGB closer to the true survival function, see \cref{fig:sfdeltaap691}.

\subsection{Sensitivities}

\begin{figure}[htbp]
    \centering 
\begin{subfigure}{0.32\linewidth}
  \includegraphics[width=\linewidth]{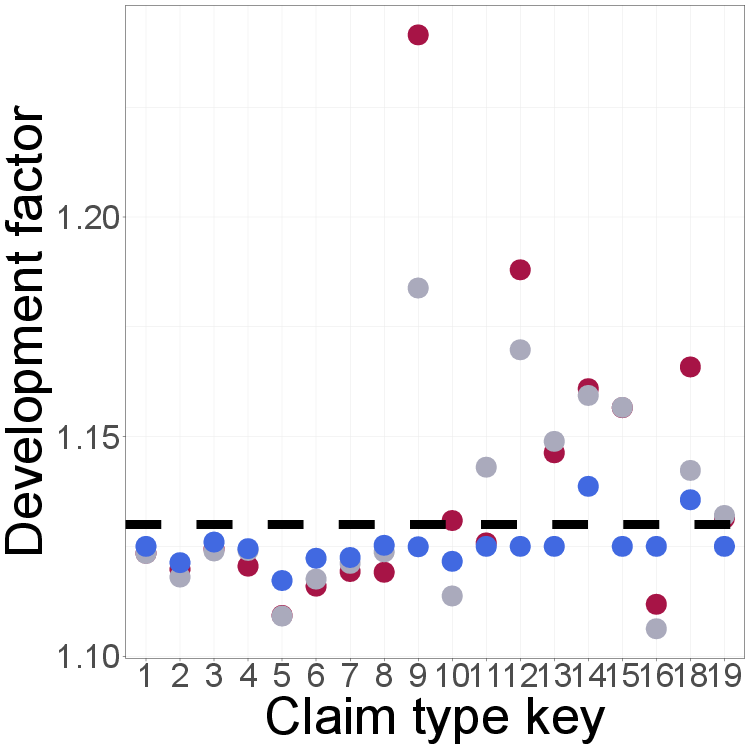}
  \caption{}
  \label{fig:a}
\end{subfigure}\hfil 
\begin{subfigure}{0.32\linewidth}
  \includegraphics[width=\linewidth]{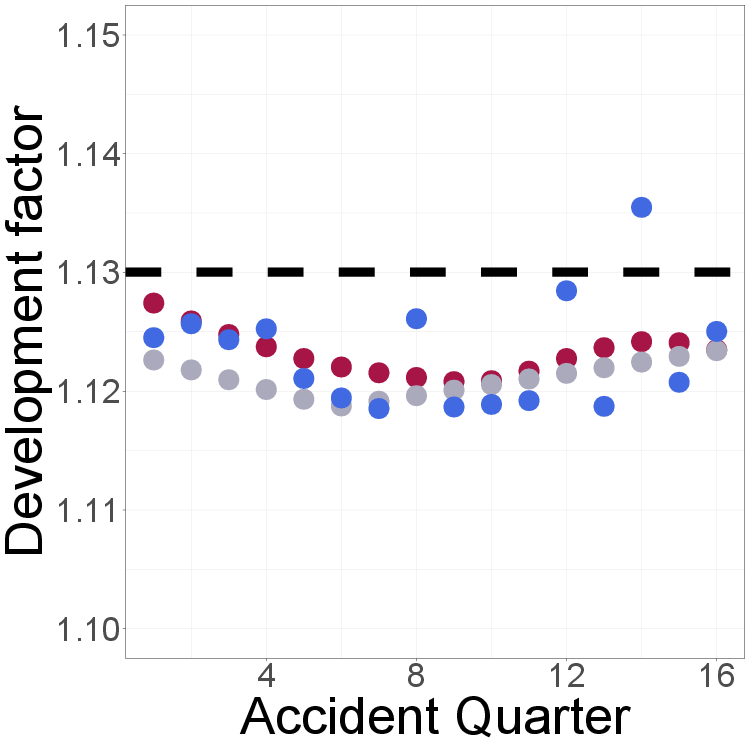}
  \caption{}
  \label{fig:b}
\end{subfigure}\hfil 
\begin{subfigure}{0.32\linewidth}
  \includegraphics[width=\linewidth]{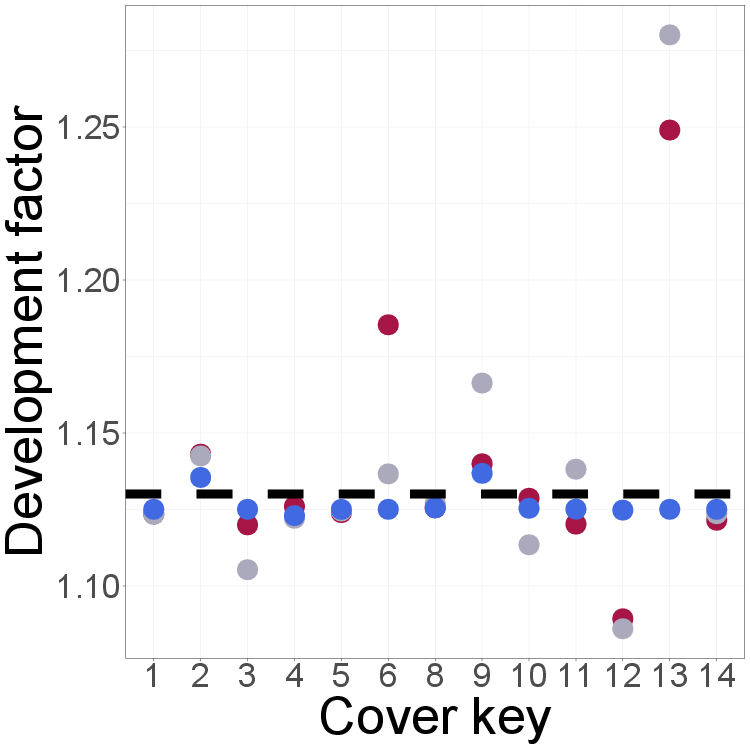}
  \caption{}
  \label{fig:c}
\end{subfigure}

\caption{\label{fig:sensitivitiesdq2} Development factor $2$ sensitivity for the \textbf{quarterly} output in Split 6. The dotted line represents the chain ladder estimate. The dots indicate the estimates from the different models: COX (red), NN (gray) and XGB (blue).}
\end{figure}

Inspired by the analysis of the sensitivities in \citeA{wuthrich18}, we show that for a fixed combination of features and a fixed development period, the marginal effect of the different levels of a model features on the development factors for Spit 1. The benchmark is the development factor modelled with the chain ladder. 
The first row shows our results on a quarterly grid (\cref{fig:a}, \cref{fig:b}, \cref{fig:c}) for the second development quarter. In \cref{fig:a} fix the accident quarter to $16$, and \texttt{cover\_key} to $1$ and let \texttt{claim\_type\_key} vary. In \cref{fig:b}, we pick \texttt{cover\_key 1}, \texttt{claim\_type\_key 1} and let the accident quarter change. In \cref{fig:c}, we pick accident quarter $16$, \texttt{claim\_type\_key 1} and let the \texttt{cover\_key} vary.

\begin{figure}[htbp]
    \centering 
\begin{subfigure}{0.32\linewidth}
  \includegraphics[width=\linewidth]{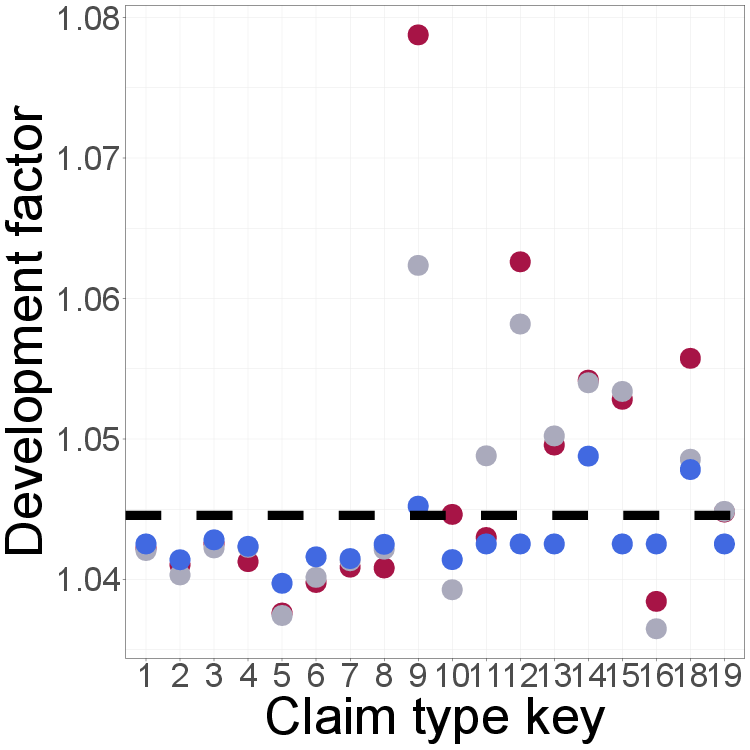}
  \caption{}
  \label{fig:d}
\end{subfigure}\hfil 
\begin{subfigure}{0.32\linewidth}
  \includegraphics[width=\linewidth]{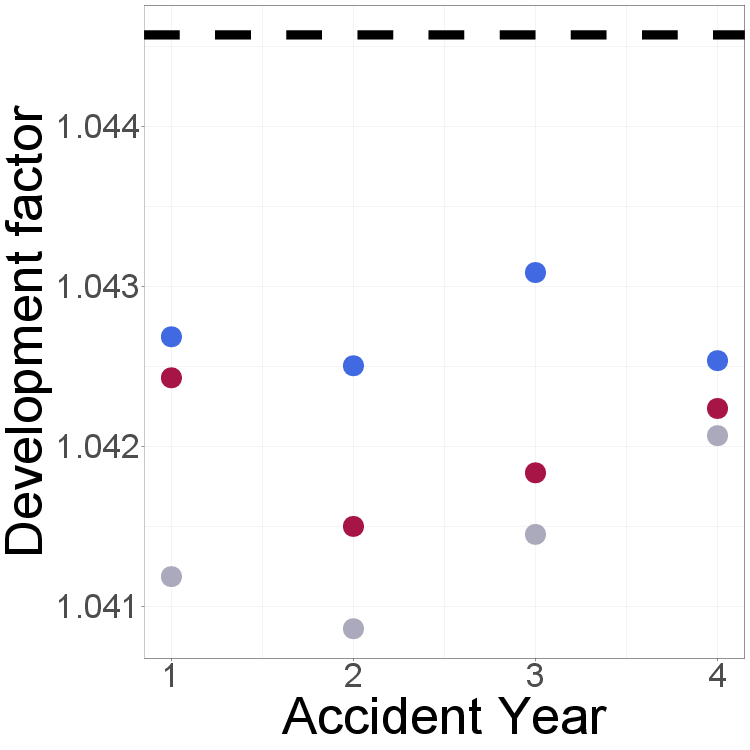}
  \caption{}
  \label{fig:e}
\end{subfigure}\hfil 
\begin{subfigure}{0.32\linewidth}
  \includegraphics[width=\linewidth]{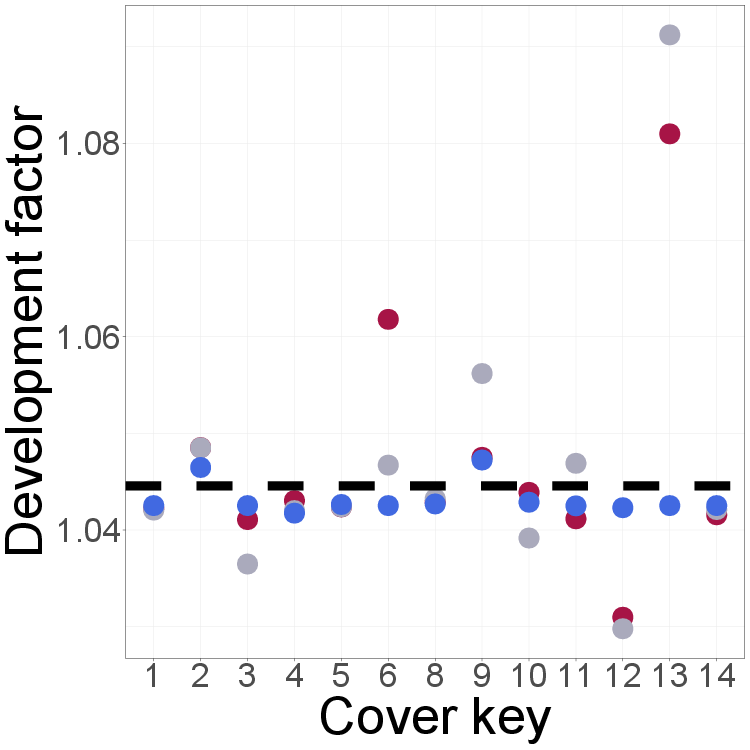}
  \caption{}
  \label{fig:f}

\end{subfigure}

\caption{\label{fig:sensitivitiesdy2} Development factor $2$ sensitivity for the \textbf{yearly} output in Split 6. The dotted line represents the chain ladder estimate. The dots indicate the estimates from the different models: COX (red), NN (gray) and XGB (blue).}
\label{fig:wuthsens}
\end{figure}

In a similar fashion, in \cref{fig:sensitivitiesdy2} we show a similar plot for the yearly results.  In \cref{fig:d} we fix the accident year to $4$, and \texttt{cover\_key} to $1$ and let \texttt{claim\_type\_key} vary. In \cref{fig:e}, we pick \texttt{cover\_key 1}, \texttt{claim\_type\_key 1} and let the accident quarter change. In \cref{fig:f}, we pick accident quarter $4$, \texttt{claim\_type\_key 1} and let the \texttt{cover\_key}.  

\section{Computational details}

The computations were performed in Linux, using
\href{https://www.erda.dk/}{ \texttt{ERDA} }(Electronic Research Data Archive, University of Copenhagen).  The relevant computational details on the architecture are provided below.

 \begin{lstlisting}
>>> lscpu

Architecture:            x86_64
CPU op-mode(s):        32-bit, 64-bit
Address sizes:         48 bits physical, 48 bits virtual
Byte Order:            Little Endian
CPU(s):                  64
On-line CPU(s) list:   0-63
Vendor ID:               AuthenticAMD
Model name:            AMD EPYC Processor (with IBPB)
CPU family:          23
Model:               1
Thread(s) per core:  1
Core(s) per socket:  1
Socket(s):           64
Stepping:            2
BogoMIPS:            4000.00
\end{lstlisting}
 
\end{document}